\documentclass[aps,prd,preprintnumbers,groupedaddress,nofootinbib,amssymb,eqsecnum,notitlepage,dvipdfmx]{revtex4}
\usepackage{here}
\usepackage{graphicx}
\usepackage{amsmath,amsthm,amssymb}
\usepackage{bm}
\usepackage{color}

\usepackage{amsfonts}
\usepackage{dcolumn}
\allowdisplaybreaks[1]

\begin{document}
\newcommand{\newc}{\newcommand}

\newcommand{\rk}[1]{{\color{red} #1}}
\newcommand{\ben}{\begin{eqnarray}}
\newcommand{\een}{\end{eqnarray}}
\newc{\be}{\begin{equation}}
\newc{\ee}{\end{equation}}
\newc{\ba}{\begin{eqnarray}}
\newc{\ea}{\end{eqnarray}}
\newc{\bea}{\begin{eqnarray*}}
\newc{\eea}{\end{eqnarray*}}
\newc{\tp}{\dot{\phi}}
\newc{\ttp}{\ddot{\phi}}
\newc{\nrhon}{n_I\rho_{I,n_I}}
\newc{\nrhocn}{n_c\rho_{c,n_c}}
\newc{\drho}{\delta \rho_I}
\newc{\drhoc}{\delta \rho_c}
\newc{\dphi}{\delta\phi}
\newc{\D}{\partial}
\newc{\ie}{{\it i.e.} }
\newc{\eg}{{\it e.g.} }
\newc{\etc}{{\it etc.} }
\newc{\etal}{{\it et al.}}
\newcommand{\nn}{\nonumber}
\newc{\ra}{\rightarrow}
\newc{\lra}{\leftrightarrow}
\newc{\lsim}{\buildrel{<}\over{\sim}}
\newc{\gsim}{\buildrel{>}\over{\sim}}
\newc{\aP}{\alpha_{\rm P}}
\newc{\delj}{\delta j}
\newc{\rhon}{\rho_{m,n}}
\newc{\rhonn}{\rho_{m,nn}}
\newc{\delrho}{\delta \rho_m}
\newc{\pa}{\partial}
\newc{\E}{{\cal E}}
\newc{\rH}{{\rm H}}

\title{Scalar-field dark energy 
nonminimally and kinetically coupled to dark matter}

\author{Ryotaro Kase and Shinji Tsujikawa}

\affiliation{Department of Physics, Faculty of Science, 
Tokyo University of Science, 1-3, Kagurazaka,
Shinjuku-ku, Tokyo 162-8601, Japan}

\begin{abstract}

We provide a general framework for studying the dark energy 
cosmology in which a scalar field $\phi$ is nonminimally and kinetically 
coupled to Cold Dark Matter (CDM). The scalar-graviton sector is 
described by the action of Horndeski theories with the 
speed of gravitational waves equivalent to that of light, 
whereas CDM is treated as a perfect fluid given by 
a Schutz-Sorkin action. We consider two  
interacting Lagrangians of the forms $f_1(\phi,X)\rho_c (n_c)$ 
and $f_2 (n_c, \phi,X) J_c^{\mu} \partial_{\mu}\phi$, where 
$X=-\partial^{\mu} \phi \partial_{\mu} \phi/2$, 
$\rho_c$ and $n_c$ are the energy density and number 
density of CDM respectively, and $J_c^{\mu}$ is 
a vector field related to the CDM four velocity. 
We derive the scalar perturbation equations of motion without 
choosing any special gauges and identify conditions for the 
absence of ghosts and Laplacian instabilities on scales 
deep inside the sound horizon. 
Applying a quasi-static approximation in a gauge-invariant manner, 
we also obtain the effective gravitational couplings felt by CDM 
and baryons for the modes relevant to the linear growth 
of large-scale structures. 
In particular, the $n_c$ dependence in the coupling $f_2$ gives 
rise to an interesting possibility for realizing  
the gravitational coupling with CDM weaker than the Newton 
gravitational constant $G$.

\end{abstract}

\date{\today}


\maketitle

\section{Introduction}
\label{introsec}

In spite of the numerous observational evidence for the existence of 
dark energy and dark matter, the origins of them have not been identified yet.
Dark energy is responsible for today's cosmic 
acceleration \cite{dreview1,dreview2,dreview3}, 
whereas dark matter is the main source for the growth of 
large-scale structures \cite{dmreview1,dmreview2}. 
The standard cosmological paradigm is 
known as the $\Lambda$-Cold Dark Matter ($\Lambda$CDM) model, 
in which dark energy and dark matter are attributed to the cosmological 
constant and weakly interacting nonrelativistic particles, 
respectively. 
The $\Lambda$CDM model is overall consistent with the current observational 
data, but there have been tensions for today's Hubble constant $H_0$ 
and the amplitude $\sigma_8$ of matter density contrast between 
the high- and low-redshift 
measurements \cite{Ade:2015xua,Riess:2016jrr,Hildebrandt:2016iqg}. 

The cosmological constant is not the only possibility for the origin of 
dark energy, but there are also other candidates like 
a time-varying scalar field $\phi$.
If we allow scalar nonminimal and derivative couplings to gravity, 
Horndeski theories \cite{Horndeski} 
are the most general scalar-tensor theories with second-order 
field equations \cite{Ho1,Ho2,Ho3}. 
Meanwhile, the gravitational wave (GW) event GW170817 \cite{GW170817} 
constrains the speed of gravity $c_t$ to be very close to that of light $c$, 
at the level $|c_t/c-1| \lesssim 10^{-15}$. 
If we strictly demand that $c_t=c$ without allowing any tuning among functions, 
the Horndeski Lagrangian is restricted to be of the form 
$L_{\rm H}=G_2(\phi,X)+G_3(\phi,X) \square \phi+G_4(\phi)R$, where  
$G_{2,3}$ are functions of $\phi$ and $X$, and $G_4$ is 
a function of $\phi$ alone multiplied 
by the Ricci scalar $R$ \cite{Lombriser:2015sxa,GW1,GW2,GW3,GW4,GW5,GW6}. 
In this theoretical scheme, there are some dark energy models 
fitting to the observational data better than 
the $\Lambda$CDM \cite{Peirone:2019aua,Peirone:2019yjs}.
Nevertheless, it is still a challenging issue to alleviate the 
tensions of $H_0$ and $\sigma_8$ between the high- and 
low-redshift measurements.

{}From the viewpoint of particle physics, it is natural to expect that
dark energy interacts with dark matter \cite{Wetterich}.
Such a coupling can potentially alleviate the coincidence problem 
as to why the energy densities of two dark components are of the 
same orders today. For a quintessence scalar field $\phi$ whose continuity 
equation is sourced by the term $\beta \rho_c \dot{\phi}$, where 
$\beta$ is coupling constant, $\rho_c$ is the CDM density, and $\dot{\phi}$ 
is the time derivative of $\phi$, there exists a scaling matter era
characterized by a constant field density parameter 
$\Omega_{\phi}=2\beta^2/3$ followed by the dark energy 
dominated epoch \cite{Amendola99a,Amendola99,Gumjudpai,Amendola06}.
In this model, the likelihood analysis with Cosmic Microwave Background (CMB) 
temperature anisotropies from {\it Planck}
and other data give rise to the upper limit $\beta<0.062$ (95\,\%\,CL), 
with a mild peak of the marginalized posterior distribution 
around $\beta=0.036$ \cite{Ade15}.

There have been also several phenomenological approaches to the interacting 
dark energy scenario in which the couplings with CDM are added to the continuity 
equations by hand \cite{Dalal:2001dt,Zimdahl:2001ar,Chimento:2003iea,Wang1,
Wei:2006ut,Amendola:2006dg,Guo:2007zk,Valiviita:2008iv,Valiviita:2009nu,Gavela:2009cy,Jackson:2009mz,Faraoni:2014vra,Wands,Wang2,DiValentino:2017iww,DiValentino:2019ffd}. 
One of such examples is to introduce the interacting term  
$\xi H \rho_{c}$ or $\xi H \rho_{\rm DE}$ 
to the right hand side of the CDM continuity equation, where $\xi$ is 
a coupling constant, $H$ is the Hubble expansion rate, and $\rho_{\rm DE}$ 
is the dark energy density. For the interaction $\xi H \rho_{\rm DE}$, the recent 
joint data analysis based on {\it Planck}, the direct measurement of $H_0$,  
galaxy clusterings, and cosmic shear measurements shows that 
the model with negative $\xi$ (i.e., the energy transfer from CDM to 
dark energy) significantly reduces the tensions of $H_0$ and $\sigma_8$ between 
CMB and low-redshift measurements below 
the 68\,\%\,CL level \cite{DiValentino:2019ffd}. 
A similar conclusion was also reached in Refs.~\cite{Berezhiani:2015yta,Chudaykin:2017ptd}
for interacting models in which a subdominant fraction of dark matter decays 
after the recombination epoch.

At the covariant level, the phenomenological interactions mentioned above 
can render difficulties for defining physical quantities properly \cite{Tamanini:2015iia}, 
e.g., the CDM velocity. 
This problem manifests itself for the dynamics of cosmological perturbations, 
leading to unphysical instabilities in the early Universe \cite{Valiviita:2008iv,Valiviita:2009nu}. 
In contrast, the Lagrangian formulation of coupled dark energy and dark matter
with their fully covariant energy-momentum tensors does not give rise to 
any ambiguities for physical quantities at the perturbation level. 
In this vein, Pourtsidou {\it et al.} \cite{Pourtsidou:2013nha} and 
Boehmer {\it et al.} \cite{Boehmer:2015kta,Boehmer:2015sha} 
constructed interacting theories of the dark sector  
by using an action of the relativistic perfect fluid for CDM 
(see also Refs.~\cite{Skordis:2015yra,Pourtsidou:2016ico,Linton:2017ged}).
The variational approach to the perfect fluid was originally advocated by 
Schutz and Sorkin \cite{Sorkin} in 1977 and was further developed 
by Brown \cite{Brown} in 1993. 
The corresponding action, which we call the Schutz-Sorkin action, contains 
an energy density $\rho$ and a vector field $J^{\mu}$ associated with 
the fluid four velocity, together with Lagrange multipliers. 
Since both scalar and vector degrees of freedom can be consistently 
implemented in this framework, the Schutz-Sorkin action has been used to describe 
perfect fluids in the late-time cosmology in scalar-tensor and vector-tensor 
theories \cite{DGS,Bettoni:2011fs,Bettoni:2015wla,Koivisto:2015qua,DeFelice:2016yws,Dutta:2016bbs,Dutta:2017wfd,Kase:2018nwt,Kase:2018aps,Frusciante:2018aew,Frusciante:2018tvu,Nakamura:2019phn}.

Neglecting the dependence of entropy density $s$, the possible interacting Lagrangian 
between the scalar field $\phi$ and the CDM density $\rho_c$ is of the form 
${\cal L}_{{\rm int}1}=-\sqrt{-g}\,f_1(\phi) \rho_c (n_c)$, where $g$ is the 
determinant of metric tensor $g_{\mu \nu}$, $f_1(\phi)$ is a function of $\phi$, 
and $\rho_c$ depends on the CDM number 
density $n_c$ \cite{Boehmer:2015kta}\footnote{
As in Refs.~\cite{Boehmer:2015kta,Frusciante:2018aew,Frusciante:2018tvu}, 
this interacting Lagrangian can be written in the more
general form ${\cal L}_{{\rm int}1}=-\sqrt{-g}\,\rho_c (n_c,\phi)$.}.
The Schutz-Sorkin Lagrangian contains the contribution 
${\cal L}_{M}=-\sqrt{-g}\,\rho_c (n_c)$, so the total effective 
CDM density is given by $\hat{\rho}_c=(1+f_1)\rho_c$. 
As we will see later, the energy transfer between CDM and the scalar field
is particularly transparent by considering continuity equations 
for $\hat{\rho}_c$ and $\rho_{\rm DE}$. 
Indeed, this interacting model accommodates 
the coupled quintessence scenario advocated in Refs.~\cite{Wetterich,Amendola99}.

Since the Schutz-Sorkin action contains a vector field $J_c^{\mu}$, 
we can also think of the scalar derivative coupling $J_c^{\mu} \partial_{\mu} \phi$. 
Indeed, the interacting Lagrangian of the form 
${\cal L}_{{\rm int}2}=f_2(n_c, \phi) J_c^{\mu} \partial_{\mu} \phi$, 
where $f_2$ is a function of  $n_c$ and $\phi$, 
was proposed in Ref.~\cite{Boehmer:2015sha}. 
This coupling opened up a new window for the study of coupled dark energy. 
For instance, the effective gravitational interaction with matter density perturbations 
can be either weaker or stronger than the Newton gravitational constant on scales 
relevant to the linear growth of large-scale structures \cite{Koivisto:2015qua}. 
This is not the case for the coupled quintessence with the interacting 
Lagrangian ${\cal L}_{{\rm int}1}$ mentioned above, in that the scalar-matter 
interaction is always attractive. Hence the signature of different 
couplings can be probed from the observations of galaxy clusterings and weak lensing. 
In particular, it will be of interest to study whether models with the weak gravitational interaction 
for the large-scale structure growth, as indicated in 
current observations \cite{Ade15,Beu,Ledo,Eriksen,Vik}, 
can be consistently constructed in this framework.

The couplings $f_1(\phi)$ and $f_2(n_c, \phi)$ can be generalized to include 
the dependence of field kinetic energy 
$X=-\partial^{\mu} \phi \partial_{\mu} \phi/2$, 
such that the two interacting Lagrangians are given by 
${\cal L}_{{\rm int}1}=-\sqrt{-g}\,f_1(\phi,X) \rho_c (n_c)$ and 
${\cal L}_{{\rm int}2}=f_2(n_c, \phi,X) J_c^{\mu} \partial_{\mu} \phi$. 
The background cosmological dynamics with the first interaction was 
recently discussed for a canonical field with the potential $V(\phi)$ \cite{Barros}.
The interacting Lagrangians ${\cal L}_{{\rm int}1}$ and ${\cal L}_{{\rm int}2}$ 
may also be partially related to the theories in which CDM is 
conformally and disformally 
coupled to the metric $\bar{g}_{\mu \nu}$ different from the metric 
$g_{\mu \nu}$ felt by baryons \cite{Gleyzes:2015pma,Kimura:2017fnq,Chibana:2019jrf}. 
So far, most of the past papers  considered the canonical scalar 
or k-essence field for the dark energy sector, but we would like to extend the analysis 
to more general coupled Horndeski theories satisfying $c_t=c$. 
This allows the possibility for realizing the dark energy equation of state 
$w_{\rm DE}$ smaller than $-1$ \cite{Peirone:2019aua,Peirone:2019yjs}. 

In this paper, we provide a general framework of coupled Horndeski scalar 
dark energy with two interacting Lagrangians ${\cal L}_{{\rm int}1}$ and 
${\cal L}_{{\rm int}2}$ containing the $X$ dependence in $f_1$ and $f_2$. 
We derive the scalar perturbation equations of motion on the  
flat Friedmann-Lema\^{i}tre-Robertson-Walker (FLRW) background 
in a gauge-ready form and obtain the stability conditions in the small-scale limit.
We also compute the effective gravitational couplings of CDM and baryons 
associated with the linear growth of large-scale structures.
This general formulation will be useful to construct theoretically consistent 
models of coupled dark energy and to confront them with observations.

This paper is organized as follows. 
In Sec.~\ref{eomsec}, we derive the background equations of motion on the 
flat FLRW spacetime and discuss how dark energy and CDM interact with 
each other.
In Sec.~\ref{actionsec}, we expand the total action up to second order 
in scalar perturbations and obtain the perturbation equations 
without fixing gauge conditions.
In Sec.~\ref{stasec}, we identify conditions for the absence of ghosts 
and Laplacian instabilities in the small-scale limit by choosing several 
different gauges and show that they are independent of the choice 
of gauges.
In Sec.~\ref{Geffsec}, we apply the quasi-static approximation to
the perturbations deep inside the sound horizon and derive the general 
expression of the effective gravitational couplings of CDM and baryons.
Sec.~\ref{consec} is devoted to conclusions.

We use the natural unit where the speed of light $c$,
the reduced Planck constant $\hbar$, and the Boltzmann 
constant $k_B$ are equivalent to 1. 
The Newton gravitational constant $G$ is related to the
reduced Planck mass $M_{\rm pl}$, as 
$G=1/(8\pi M_{\rm pl}^2)$.
The Greek and Latin indices denote components in space-time 
and in a three-dimensional space-adapted basis, respectively.

\section{Lagrangian formulation of coupled dark energy}
\label{eomsec}

We consider the Lagrangian formulation of scalar-field dark energy 
coupled to CDM. The scalar-graviton sector is described by 
the action of Horndeski theories ${\cal S}_{\rm H}$ with the 
tensor propagation speed $c_t$ equivalent to $1$. 
Besides CDM, we also take baryons and radiations into account 
and assume that they do not have direct couplings to 
the scalar field $\phi$. 
The noninteracting matter sector of perfect fluids is described 
by the Schutz-Sorkin action ${\cal S}_M$ \cite{Sorkin,Brown}. 
For the coupled action ${\cal S}_{\rm int}$ between CDM 
and $\phi$, we consider the two types of interactions 
mentioned in Introduction. 
Then, the total action is given by 
\be
{\cal S} = {\cal S}_{\rH}
+{\cal S}_{M}
+{\cal S}_{\rm int}\,,
\label{action}
\ee
where
\ba
{\cal S}_{\rH} &=& 
\int {\rm d}^4 x \sqrt{-g} 
\left[ G_4(\phi)R+G_2(\phi,X)+G_3(\phi,X) \square \phi
\right]\,,\label{Sg}\\
{\cal S}_{M} &=&  -\sum_{I=c,b,r}\int {\rm d}^{4}x \left[
\sqrt{-g}\,\rho_I(n_I)
+ J_I^{\mu} \partial_{\mu} \ell_I \right]\,, 
\label{SM}\\
{\cal S}_{\rm int} &=& 
-\int {\rm d}^4x \sqrt{-g}\,f_1(\phi, X)\,\rho_{c} (n_c)
+\int {\rm d}^4x\,f_2(n_c, \phi, X) J_c^{\mu} \partial_{\mu} \phi\,.
\label{Sint}
\ea
Here, $R$ is the Ricci scalar, 
$G_4$ is a function of $\phi$ alone, 
$G_{2,3}$ depend on both $\phi$ and 
$X=-\partial^{\mu}\phi \partial_{\mu} \phi/2$, and 
$\square =g^{\mu \nu} \nabla_{\mu} \nabla_{\nu}$
is the d'Alembertian with the covariant derivative operator 
$\nabla_{\mu}$. The Schutz-Sorkin action (\ref{SM})
describes the perfect fluids of CDM, baryons, and 
radiations, labeled by $c,b,r$, respectively.
The energy density $\rho_I$ is a function of 
each fluid number density $n_I$, which can be 
expressed in terms of the vector field $J_I^{\mu}$, as
\be
n_I=\sqrt{\frac{J_I^{\mu} J_I^{\nu}
g_{\mu \nu}}{g}}\,.
\label{ndef}
\ee
The fluid four-velocity $u_{I{\mu}}$ is defined by 
\be
u_{I{\mu}} \equiv \frac{J_{I{\mu}}}{n_I\sqrt{-g}}\,.
\label{udef}
\ee
{}From Eq.~(\ref{ndef}), there is the relation 
$u_I^{\mu} u_{I{\mu}}=-1$. 
The scalar quantity $\ell_I$ in ${\cal S}_M$ is a Lagrange multiplier, 
with the notation of the partial derivative 
$\partial_{\mu} \ell_I \equiv \partial \ell_I/\partial x^{\mu}$. 
Throughout the paper, we do not include the contribution of 
entropy density $s_I$ per particle \cite{Sorkin,Brown} to the matter acton.
The vector degrees of freedom are generally present 
in ${\cal S}_M$ \cite{DeFelice:2016yws,Kase:2018nwt}, 
but we do not take them into account as they are 
irrelevant to the cosmological dynamics in scalar-tensor theories.

Now, we are considering interacting theories in which CDM is coupled 
to the scalar field through the dependence of both $\phi$ and $X$. 
The scalar quantities constructed from the one derivative 
$\partial_{\mu} \phi$ correspond to 
$X=-\partial^{\mu} \phi\partial_{\mu} \phi/2$ and 
$Y=J^{\mu}_c \partial_{\mu} \phi$. 
The first interacting action in Eq.~(\ref{Sint}) 
possesses the $\phi$ and $X$ dependent coupling $f_1(\phi,X)$ 
with the CDM density $\rho_c(n_c)$. 
The second interacting action in Eq.~(\ref{Sint})  
contains the derivative coupling $Y$ as the form 
$f_2(n_c, \phi, X)Y$, where $f_2$ is a general function of 
$n_c$, $\phi$, and $X$.

The theories in Ref.~\cite{Boehmer:2015kta}
can be recovered by choosing the couplings $f_1=f_1(\phi)$ and 
$f_2=0$ with $G_4=M_{\rm pl}^2/2$, $G_2=X-V(\phi)$, 
and $G_3=0$, whereas the interactions in Ref.~\cite{Boehmer:2015sha}
correspond to $f_1=0$ and $f_2=f_2(n_c,\phi)$.
Our analysis encompasses these two coupled dark energy theories 
as special cases. We note that the analysis can be extended to 
more general couplings\footnote{We thank Edmund Copeland for pointing out 
this issue after the initial submission of our paper on arXiv.} 
containing the nonlinear terms in 
$Y$, e.g., $f_2(n_c,\phi,X,Y)$ \cite{Pourtsidou:2013nha,Skordis:2015yra,Pourtsidou:2016ico,Linton:2017ged}.
The theories in which CDM is conformally and disformally coupled to the metric 
$\bar{g}_{\mu\nu}=A(\phi,X)g_{\mu\nu}+B(\phi,X)\partial_{\mu}\phi\partial_{\nu}\phi$ \cite{Gleyzes:2015pma,Kimura:2017fnq,Chibana:2019jrf} 
may be accommodated in this extended class since the term $J^{\mu}_cJ^{\nu}_c\bar{g}_{\mu\nu}$ in the definition of Eq.~(\ref{ndef}) generates 
a nonlinear term in $Y$. 
In this paper we focus on the interacting theories containing the linear term 
in $Y$, leaving more general couplings for a future work.

Variation of the action (\ref{action}) with respect to $\ell_I$ leads to
\be
\partial_{\mu} J_{I}^{\mu}=0\,,\qquad 
{\rm for} \qquad I=c,b,r\,.
\label{Jmure}
\ee
Varying the action (\ref{action}) with respect to $J_I^{\mu}$ 
and using the relation $\partial n_I/\partial J_I^{\mu}
=J_{I{\mu}}/(n_I g)$, it follows that 
\ba
& &
\partial_{\mu} \ell_{c}= 
u_{c{\mu}} \rho_{c,n_c} \left( 1+f_1 \right) 
-\frac{1}{\sqrt{-g}} u_{c \mu} f_{2,n_c}
J_c^{\alpha} \partial_{\alpha}\phi
+f_2 \partial_{\mu} \phi\,,
\label{lc}
\\
& &
\partial_{\mu} \ell_{I}
= u_{I{\mu}} \rho_{I,n_I}\,, \qquad \qquad 
\qquad \,{\rm for}~I=b,r\,,
\label{l1}
\ea
where $\rho_{I,n_I} \equiv \partial \rho_I/\partial n_I$
and  $f_{2,n_c} \equiv \partial f_2/\partial n_c$.

\subsection{Background equations of motion}

We derive the field equations of motion on the flat FLRW 
background given by the line element
\be
{\rm d} s^{2} = - N^2(t) {\rm d}t^2
+a^2(t) \delta_{ij} {\rm d}x^{i} {\rm d}x^{j}\,,
\label{metric}
\ee
where $N(t)$ is the lapse, $a(t)$ is the scale factor, 
and $t$ is the cosmic time. 
Since the fluid four-velocity in its rest frame is given by 
$u_I^{\mu}=(N^{-1},0,0,0)$, Eq.~(\ref{udef}) gives 
the relation $J_I^{0}=n_I \sqrt{-g}\, u_I^{0}$, i.e., 
\be
J_I^0=n_I a^3\,.
\label{nJ}
\ee
{}From Eq.~(\ref{Jmure}), we obtain
\be
{\cal N}_I \equiv J_I^0=n_I a^3={\rm constant}\,,
\label{Ji}
\ee
which means that the particle number ${\cal N}_I$ of 
each matter species is conserved.

On the background (\ref{metric}), the action (\ref{action}) reduces to 
\ba
{\cal S}
&=&\int {\rm d}^4 x \left\{ \frac{a^3}{N} \left[ N^2 G_2(\phi, X)
-\left( \ddot{\phi}+3\frac{\dot{a}}{a} \dot{\phi} \right)G_3(\phi,X)
+6 \left( \frac{\ddot{a}}{a}+\frac{\dot{a}^2}{a^2} 
\right) G_4 (\phi) \right]+\frac{\dot{N}}{N^2}a^3
\left[ \dot{\phi}\,G_3(\phi,X)-6\frac{\dot{a}}{a} G_4(\phi) 
\right] \right\} \nonumber \\
& &-\int {\rm d}^4 x \left[ N a^3 \left\{ [1+f_1(\phi,X)]\rho_c
+\rho_b+\rho_r \right\}+a^3 \left( n_c \dot{\ell}_c 
+ n_b \dot{\ell}_b+ n_r \dot{\ell}_r \right)
-n_c a^3 f_2 (n_c, \phi, X) \dot{\phi} \right]\,, 
\label{action2}
\ea
where a dot represents the derivative with respect to $t$. 
The field kinetic energy $X$ is given by 
\be
X=\frac{\dot{\phi}^2}{2N^2}\,,
\ee
which contains the $N$ dependence.

{}From Eq.~(\ref{Ji}), the number densities $n_I$ 
satisfy the differential equations $\dot{n}_I+3Hn_I=0$ (with $I=c,b,r$), 
where $H\equiv \dot{a}/a$ is the Hubble expansion rate. 
The pressure of each matter component is given by 
\be
P_{I}(n_I)=n_I \rho_{I,n_I}-\rho_I\,.
\ee
In terms of $\rho_I(n_I)$ and $P_I(n_I)$, the conservations 
of particle numbers translate to the continuity equations,
\be
\dot{\rho}_I+3H \left( \rho_I+P_I 
\right)=0\,.
\label{coneq}
\ee
In what follows, we will consider the perfect fluid satisfying 
the weak energy conditions, i.e., 
\be
\rho_I>0\,,\qquad \rho_I+P_I>0\,.
\label{weak}
\ee
On the background (\ref{metric}), Eqs.~(\ref{lc}) 
and (\ref{l1}) reduce, respectively, to 
\ba
\dot{\ell}_c &=& -N\rho_{c,n_c} 
\left( 1+f_1 \right)+\left( n_c f_{2,n_c}
+f_2\right) \dot{\phi}\,,
\label{lre1}\\
\dot{\ell}_I &=& -N\rho_{I,n_I}\,,
\qquad \qquad \qquad\,
{\rm for}~I=b,r\,.\label{lre2}
\ea

Varying the action (\ref{action}) with respect to $N$, $a$, $\phi$, 
using Eqs.~(\ref{lre1})-(\ref{lre2}), 
and setting $N=1$ at the end, 
we obtain the background equations,
\ba
& & 
{\cal C}_1+\dot{\phi}^2 f_{1,X}\rho_c
-n_c \dot{\phi}^3 f_{2,X}-\left( 1+f_1  \right) \rho_{c} 
-\rho_{b}-\rho_{r}=0 \,,
\label{Eq00}\\
& & 
{\cal C}_2-n_c^2 f_{2,n_c}  \dot{\phi}  
+\left( 1+f_1  \right) P_{c}+P_b+P_r=0\,,
\label{Eq11}\\
& &
{\cal C}_3+\left[ f_{1,\phi}-\dot{\phi}^2 f_{1,X \phi} 
-\ddot{\phi} \left( f_{1,X}+\dot{\phi}^2f_{1,XX}  \right) 
\right] \rho_c+3H \dot{\phi} f_{1,X}P_c 
\nonumber \\
& &
-n_c \left( 3H n_c f_{2,n_c}-3\dot{\phi} \ddot{\phi}f_{2,X} 
+3H \dot{\phi}^2 n_c f_{2,X n_c} -\dot{\phi}^3 f_{2,X \phi}
-\dot{\phi^3} \ddot{\phi} f_{2,XX}
\right)
=0\,,
\label{Eqphi}
\ea
where 
\ba
{\cal C}_1&=& 
6G_4 H^{2}+ G_{2}-\dot{\phi}^2 G_{2,X}
-\dot{\phi}^2 \left( G_{3,\phi}-3H \dot{\phi} G_{3,X} 
\right)+6H \dot{\phi} G_{4,\phi}\,,\label{C1}\\
{\cal C}_2 &=& 
{\cal C}_1+2q_t \dot{H}-D_6 \ddot{\phi}+D_7 \dot{\phi}\,,\\
{\cal C}_3 &=& 
2D_1 \ddot{\phi}+3D_6 \dot{H}-D_5+3H D_7\,,
\label{C3}
\ea
and
\ba
q_t &=& 2G_4\,,
\label{qt} \\
D_1 &=& \frac{1}{2} G_{2,X}+G_{3,\phi}+\frac{1}{2} \dot{\phi}^2 
\left( G_{2,XX}+G_{3,X \phi} \right) 
-\frac{3}{2} H \dot{\phi} \left( 2 G_{3,X} +\dot{\phi}^2 
G_{3,XX} \right)\,,\label{D1}\\
D_5 &=& G_{2,\phi}-\dot{\phi}^2 \left( G_{2,X \phi}+G_{3,\phi \phi} 
\right)+3H \dot{\phi} \left( \dot{\phi}^2 G_{3,X \phi}
+2G_{4, \phi \phi} \right)+6H^2 G_{4, \phi}\,,\\
D_6 &=& -\dot{\phi}^2 G_{3,X}-2G_{4, \phi}\,,\\
D_7 &=& \dot{\phi} \left( G_{2,X}+2G_{3,\phi}+2G_{4, \phi \phi} 
\right)-H \left( 3 \dot{\phi}^2 G_{3, X}+2G_{4, \phi} \right)\,.
\label{D7}
\ea
As we will see in Sec.~\ref{tensec}, the quantity $q_t$ is associated 
with the no-ghost condition of tensor perturbations. 
The coefficients $D_{1,5,6,7}$ appear in the second-order action of 
scalar perturbations derived later in Sec.~\ref{actionsec}. 

As long as the condition 
\be
q_s \equiv 4D_1 q_t+3D_6^2
-2q_t \left( f_{1,X}+\dot{\phi}^2 
f_{1,XX} \right) \rho_c 
+2q_t n_c \dot{\phi} \left(
3f_{2,X}+\dot{\phi}^2 f_{2,XX} \right)
\neq 0 
\label{qs}
\ee
is satisfied, Eqs.~(\ref{Eq11}) and (\ref{Eqphi}) can be solved 
for $\dot{H}$ and $\ddot{\phi}$.
As we will show in Sec.~\ref{stasec}, the absence of scalar ghosts requires that 
$q_s>0$, under which the background system does not cross 
the singular point at which the corresponding determinant 
vanishes (i.e., $q_s=0$).

We define the CDM density $\hat{\rho}_c$ and the pressure 
$\hat{P}_c$ containing the effect of 
interactions with the scalar field, as
\ba
\hat{\rho}_c &\equiv& 
\left( 1+f_1 \right) \rho_{c}\,,\label{trhoc}\\
\hat{P}_c &\equiv& 
\left( 1+f_1 \right) P_{c}\,,\label{Prhoc}
\ea
together with the CDM effective equation of state, 
\be
\hat{w}_c \equiv \frac{\hat{P}_c}{\hat{\rho}_c}
=\frac{P_c}{\rho_c}\,.
\label{twc}
\ee
The coupling $f_1$ modifies both the energy density 
and pressure of CDM. For the pressureless dust ($P_c=0$), 
the effective pressure $\hat{P}_c$ vanishes.

On using the continuity Eq.~(\ref{coneq}), 
the energy density $\hat{\rho}_c$ obeys
\be
\dot{\hat{\rho}}_c+3H \left( \hat{\rho}_c
+\hat{P}_c \right)
=\frac{f_{1,\phi}+\ddot{\phi} f_{1,X}}{1+f_1} 
\hat{\rho}_c \dot{\phi}\,.
\label{rhoccon}
\ee
We can express Eqs.~(\ref{Eq00}) and (\ref{Eq11}) 
in the following forms,
\ba
3M_{\rm pl}^2 H^2&=& \rho_{\rm DE}+\hat{\rho}_c
+\rho_b+\rho_r \,,\label{back1}\\
M_{\rm pl}^2 \left( 2 \dot{H}+ 3H^{2} \right)
&=& -P_{\rm DE}-\hat{P}_c-P_b-P_r\,,
\ea
where 
\ba
\hspace{-0.5cm}
\rho_{\rm DE} &=&  - G_{2}+\dot{\phi}^2 G_{2,X}
+\dot{\phi}^2 \left( G_{3,\phi}-3H \dot{\phi} G_{3,X} 
\right)-6H \dot{\phi} G_{4,\phi}
+ 3\left(M_{\rm pl}^2-2G_4 \right)H^2
-\dot{\phi}^2 f_{1,X}\rho_c+n_c \dot{\phi}^3 f_{2,X}\,,
\label{rhode}\\
\hspace{-0.5cm}
P_{\rm DE} &=& G_{2}+\dot{\phi}^2 \left( G_{3,\phi}
+\ddot{\phi} G_{3,X} \right) 
+2G_{4,\phi} \left( \ddot{\phi}+2H \dot{\phi} \right) 
+2\dot{\phi}^2 G_{4,\phi \phi}
-\left( 2\dot{H}+3H^2 \right) \left( M_{\rm pl}^2-2G_4 
\right)-n_c^2 f_{2,n_c} \dot{\phi}\,.
\label{Pde}
\ea
Differentiating Eq.~(\ref{rhode}) with respect to $t$ and 
exploiting Eq.~(\ref{Eqphi}), we obtain
\be
\dot{\rho}_{\rm DE}+3H \left( \rho_{\rm DE}
+P_{\rm DE} \right)=-\frac{f_{1,\phi}+\ddot{\phi} f_{1,X}}{1+f_1} 
\hat{\rho}_c \dot{\phi} \,.
\label{rhodecon}
\ee
The sign on the right hand side of Eq.~(\ref{rhodecon}) is 
opposite to that of Eq.~(\ref{rhoccon}), showing the energy 
exchange between the scalar field and CDM.  
For the couplings $f_1(\phi)=e^{Q\phi/M_{\rm pl}}-1$ and $f_2=0$, 
where $Q$ is a dimensionless constant, the right hand sides of Eqs.~(\ref{rhoccon}) 
and (\ref{rhodecon}) reduce, respectively, to 
$Q \hat{\rho}_c \dot{\phi}/M_{\rm pl}$ and 
$-Q \hat{\rho}_c \dot{\phi}/M_{\rm pl}$. 
This corresponds to the coupled dark energy scenario 
originally proposed in Refs.~\cite{Wetterich,Amendola99}.

The coupling $f_2(n_c,\phi,X)$ does not give rise to its contribution to the 
right hand sides of Eqs.~(\ref{rhoccon}) and (\ref{rhodecon}).
This reflects the fact that the interaction induced by 
$f_2(n_c,\phi,X)$ corresponds to the momentum 
transfer \cite{Pourtsidou:2013nha,Skordis:2015yra,Pourtsidou:2016ico}. 
By defining
$\bar{P}_c=\hat{P}_c-n_c^2 f_{2,n_c} \dot{\phi}$ and 
$\bar{P}_{\rm DE}=P_{\rm DE}+n_c^2 f_{2,n_c} \dot{\phi}$, 
respectively, the right hand sides of continuity equations 
for CDM and dark energy acquire the terms 
$-3Hn_c^2 f_{2,n_c} \dot{\phi}$ and 
$3Hn_c^2 f_{2,n_c} \dot{\phi}$, respectively. 
Here, we do not choose these definitions of effective pressures 
to show explicitly that the energy transfer solely comes from 
the coupling $f_1(\phi,X)$. 
With $\hat{\rho}_c$ and $\hat{P}_c$ defined by
Eqs.~(\ref{trhoc}) and (\ref{Prhoc}), the CDM 
effective equation of state (\ref{twc}) is also equivalent to 
the standard value $P_c/\rho_c$.

\subsection{Tensor perturbations}
\label{tensec}

We consider the propagation of tensor perturbations $h_{ij}$ 
obeying the traceless and transverse conditions $h _{i}^{i}=0$ 
and $\partial _{i}h_{ij}=0$ on the flat FLRW background. 
The perturbed line element containing $h_{ij}$ is given by
\be
{\rm d}s^2=-{\rm d}t^2+a^2(t)(\delta_{ij}+h_{ij})\,
{\rm d}x^i {\rm d}x^j\,.
\ee
The nonvanishing components of $h_{ij}$ can be chosen 
as $h_{11}=h_1(t,z)$, $h_{22}=-h_1(t,z)$, and 
$h_{12}=h_{21}=h_2(t,z)$, where 
the two independent modes $h_1$ and $h_2$ 
depend on $t$ and $z$.
We expand the total action (\ref{action}) up to second order 
in $h_1, h_2$ and use the background Eqs.~(\ref{Eq11}) 
and (\ref{Eqphi}). After the integration by parts, the resulting 
second-order tensor action yields 
\be
{\cal S}_t^{(2)}=\int {\rm d}t\,{\rm d}^3x 
\sum_{i=1}^{2}
\frac{a^3}{4}q_t \left[ \dot{h}_i^2-\frac{c_t^2}{a^2} 
(\partial h_i)^2 \right]\,,
\label{actionSt}
\ee
where $q_t$ is defined by Eq.~(\ref{qt}), and 
\be
c_t^2=1\,.
\ee
The tensor ghost is absent under the condition 
\be
q_t=2G_4>0\,.
\ee
Since the propagation speed $c_t$ of tensor perturbations is equivalent to 
that of light, the coupled dark energy theory given by the action (\ref{action}) 
is consistent with the observational bound of $c_t$ constrained 
from the GW170817 event \cite{GW170817}. 
The values of $q_t$ and $c_t^2$ are equivalent to those 
in the uncoupled theories ($f_1=0$ and $f_2=0$), 
so the interaction between the scalar field and CDM does not affect 
the propagation of gravitational waves.

\section{Scalar perturbation equations in gauge-ready form}
\label{actionsec}

In this section, we first expand the action (\ref{action}) up to second order 
in scalar perturbations without choosing any particular gauges. 
Then, we derive linear perturbation equations of motion in the form 
ready for fixing any gauge conditions, i.e., the 
gauge-ready form \cite{Hwang:2001qk,Kase:2018aps,Heisenberg:2018wye}.

The general perturbed line element containing four scalar metric 
perturbations $\alpha, \chi, \zeta, E$ 
is given by \cite{Bardeen,Kodama:1985bj,Mukhanov:1990me}
\be
{\rm d}s^2=-(1+2\alpha) {\rm d}t^2
+2 \partial_i \chi {\rm d}t {\rm d}x^i
+a^2(t) \left[ (1+2\zeta) \delta_{ij}
+2\partial_i \partial_j E \right] {\rm d}x^i {\rm d}x^j\,,
\label{permet}
\ee
where the perturbed quantities depend on $t$ and $x^i$. 
The scalar field $\phi$ is also decomposed as 
\be
\phi=\bar{\phi}(t)+\delta \phi\,,
\ee
where $\bar{\phi}(t)$ is the background value and 
$\delta \phi$ is the perturbation.
In the following, we omit the overbar from background quantities.  
The perturbation of field kinetic energy $X$, expanded up to 
second order, is given by 
\be
\delta X=\dot{\phi} (\dot{\delta \phi}-\dot{\phi} \alpha)
+\frac{1}{2} \left[  (\dot{\delta \phi}-2\dot{\phi} \alpha)^2
-\frac{1}{a^2}(\partial \delta \phi+\dot{\phi} \partial \chi)^2
\right]+{\cal O} (\varepsilon^3)\,,
\label{deltaX}
\ee
where $\varepsilon^n$ represents the $n$-th order of perturbations.

For quantities in the Schutz-Sorkin action (\ref{SM}), we decompose 
the temporal and spatial components of 
$J_I^{\mu}$ (with $I=c,b,r$) 
into the background and perturbed parts, as 
\be
J_I^{0} = \mathcal{N}_{I}+\delta J_I\,,\qquad
J_I^{i} =\frac{\delta^{ik}\partial_{k}\delta j_I}{a^2(t)}\,,
\label{elldef}
\ee
where $\mathcal{N}_{I}$ is the conserved particle number, and 
$\delta J_I, \delta j_I$ correspond to scalar perturbations. 
For baryons and radiations, we recall that the Lagrange multipliers 
$\ell_I$ satisfy the relations (\ref{l1}). 
The velocity potentials $v_I$ are defined by 
\be
u_{Ii}=-\partial_{i}v_I\,.
\ee
Since $J_{Ii}=J_I^0 g_{0i}+J_I^j g_{ij}={\cal N}_I \partial_{i}\chi+
\partial_i \delta j_I$ at linear order in perturbations, 
Eq.~(\ref{udef}) gives 
\be
\partial_i \delta j_{I} 
= -\mathcal{N}_{I} \left( \partial_i \chi + \partial_i v_{I} \right) \,.
\label{deljre}
\ee
{}From Eq.~(\ref{lc}), the spatial derivative of $\ell_c$ 
associated with CDM perturbations yields 
\be
\partial_i \ell_c=-\left\{ \rho_{c,n_c}(t) [1+f_1(t)]
-n_c(t) f_{2,n_c}(t) \dot{\phi}(t) \right\} \partial_i v_c
+f_2(t) \partial_{i} \delta \phi\,, 
\ee
whose integrated solution is
\be
\ell_c={\cal A}(t)-\left\{ \rho_{c,n_c}(t) [1+f_1(t)]-n_c(t) 
f_{2,n_c}(t) \dot{\phi}(t) \right\} v_c
+f_2(t) \delta \phi\,.
\ee
The time-dependent function ${\cal A}(t)$ is determined by 
the condition that $\partial_{0} \ell_c$ computed from Eq.~(\ref{lc}) 
coincides with $\dot{\cal A}(t)$ at the background level. 
Then, it follows that 
\be
\ell_c
=-\int^t \left\{ \rho_{c,{n_c}}(\tilde{t})\left[ 1+f_1(\tilde{t}) \right]
-[n_c (\tilde{t})f_{2,n_c}(\tilde{t})+f_2 (\tilde{t})]
\dot{\phi}(\tilde{t}) 
\right\}
\,{\rm d}\tilde{t}
-\left\{ \rho_{c,n_c}(t) [1+f_1(t)]-n_c(t) 
f_{2,n_c}(t) \dot{\phi}(t) \right\} v_c
+f_2(t) \delta \phi.
\label{ellc}
\ee
For baryons and radiations, Eq.~(\ref{l1}) gives
\be
\ell_I
=-\int^t \rho_{I,{n_I}}(\tilde{t})\,{\rm d}\tilde{t}
-\rho_{I,{n_I}}(t)\,v_I\,,\qquad 
{\rm for}~I=b,r\,.
\ee

We define the density perturbation of each matter fluid, as
\be
\delta \rho_I \equiv \frac{\rho_{I,n_I}}{a^3} 
\left[ \delta J_I-{\cal N}_I \left( 3\zeta+\partial^2 E 
\right) \right]\,.
\ee
Then, the perturbation of $n_I$, which is expanded up to 
second order, is given by 
\be
\delta n_I= \frac{\delta \rho_I}{\rho_{I,n_I}}
-\frac{{\cal N}_I (\partial v_I)^2}{2a^5}
-(3\zeta+\partial^2E)\frac{\delta \rho_I}{\rho_{I,n_I}}
-\frac{{\cal N}_I(\zeta+\partial^2E)(3\zeta-\partial^2E)}{2a^3}
+{\cal O}(\varepsilon^3)\,,
\label{deln}
\ee
whose first term on the right hand side shows the consistency 
with the left hand side. 

The fluid density $\rho_I$, which depends on $n_I$, is 
expressed in the form  
\ba
\rho_I  (n_I)
&=&\rho_I +\rho_{I,n_I} \delta n_I
+\frac{1}{2} \rho_{I,n_I n_I} \delta n_I^2
+{\cal O} (\varepsilon^3) \nonumber \\
&=& \rho_I+\left( \rho_I+P_I \right) 
\frac{\delta n_I}{n_I}+\frac{1}{2} \left( \rho_I+P_I \right) 
c_I^2 \left( \frac{\delta n_I}{n_I} \right)^2
+{\cal O} (\varepsilon^3) \,,
\ea
where $c_I^2$ is the matter sound speed squared defined by 
\be
c_I^2 =\frac{P_{I,n_I}}{\rho_{I,n_I}}
=\frac{n_I \rho_{I,n_I n_I}}{\rho_{I,n_I}}\,. 
\label{ccbr}
\ee
In the interacting action (\ref{Sint}), 
the coupling terms $f_1$ and $f_2$ are also expanded as 
\ba
f_1(\phi,X) &=&
f_1+f_{1,\phi} \delta \phi+f_{1,X} \delta X
+\frac{1}{2} f_{1,\phi \phi} \delta \phi^2+
\frac{1}{2} f_{1,XX} \delta X^2
+f_{1,X \phi} \delta X\delta \phi +{\cal O} (\varepsilon^3)\,,\\
f_2(n_c,\phi,X) &=&
f_2+f_{2,n_c} \delta n_c+f_{2,\phi} \delta \phi+f_{2,X} \delta X
+\frac{1}{2} f_{2,n_c n_c} \delta n_c^2
+\frac{1}{2} f_{2,\phi \phi} \delta \phi^2
+\frac{1}{2} f_{2,XX} \delta X^2 \nonumber \\
& & +f_{2,n_c \phi} \delta n_c \delta \phi
+f_{2,n_c X} \delta n_c \delta X
+f_{2,X \phi} \delta X \delta \phi
+{\cal O} (\varepsilon^3)\,,
\ea
where $\delta X$ and $\delta n_c$ are given, respectively, 
by Eqs.~(\ref{deltaX}) and (\ref{deln}).

\subsection{Second-order action}

We first expand the Horndeski action (\ref{Sg}) 
up to quadratic order in scalar perturbations without using the 
background Eqs.~(\ref{Eq00})-(\ref{Eqphi}). 
After the integration by parts, the second-order 
Horndeski action yields
\be
{\cal S}_{\rH}^{(2)}=\int 
{\rm d}t\,{\rm d}^3x \left( L_{\rH}^{\rm flat}
+L_{\rH}^{\zeta}+L_{\rH}^{E} \right)\,,
\label{SH2}
\ee
where 
\ba
L_{\rH}^{\rm flat}
&=& a^3\left[
D_1\dot{\dphi}^2+D_2\frac{(\D\dphi)^2}{a^2}+D_3\dphi^2
+\left(D_4\dot{\dphi}+D_5\dphi+D_6\frac{\D^2\dphi}{a^2}\right) \alpha
-\left(D_6\dot{\dphi}-D_7\dphi\right)\frac{\D^2\chi}{a^2}
\right.\notag\\
&&\left.~~~
+\left(\dot{\phi} D_6-2Hq_t\right)\alpha\frac{\D^2\chi}{a^2}
+\left(\dot{\phi}^2 D_1+3H \dot{\phi} D_6-3H^2 q_t\right)\alpha^2
+\frac{1}{2}{\cal C}_1\left\{\frac{(\D\chi)^2}{a^2}-\alpha^2\right\}
\right] \,,
\label{Lgf}\\
L_{\rH}^{\zeta}
&=& a^3\left[
\left\{
3D_6\dot{\dphi}-3D_7\dphi-3\left(\dot{\phi} D_6-2Hq_t\right)\alpha
+2q_t\frac{\D^2\chi}{a^2}\right\}\dot{\zeta}-3q_t\dot{\zeta}^2
+q_t\frac{(\D\zeta)^2}{a^2}
\right.\notag\\
&&\left.~~~
-\left(B_1\dphi+2q_t\alpha\right)\frac{\D^2\zeta}{a^2}
+3\left({\cal C}_1\alpha+\frac12{\cal C}_2\zeta-{\cal C}_3\dphi\right)\zeta
\right]\,,
\label{Lgze}\\
L_{\rH}^{E}
&=& a^3\left[
2q_t\ddot{\zeta}+2B_2\dot{\zeta}
-D_6\ddot{\dphi}-B_3\dot{\dphi}+B_4\dphi
+\frac{1}{a^3}\frac{{\rm d}}{{\rm d}t}\left\{a^3 \left(\dot{\phi} D_6-2Hq_t 
\right) \alpha\right\}\right.\notag\\
&&\hspace{0.7cm}
\left.
+{\cal C}_1\alpha+{\cal C}_2\left(\zeta-\frac12\D^2E\right)
-{\cal C}_3\dphi
\right]\,\D^2 E\,,
\label{LgE}
\ea
with the coefficients
\ba
&&
D_2 = -\frac12 G_{2,X}-G_{3,\phi}+2 H{\tp} G_{3,X}
+\frac12 {\tp}^{2}G_{3,X\phi}
+\frac12\left( 2G_{3,X}+G_{3,{XX}}{\tp}^{2} \right) {\ttp}
\,,\notag\\
&&
D_3 =\frac12 G_{{2,\phi\phi}} 
-\frac12  \left(  
G_{{2,X\phi\phi}}+  G_{{3,\phi\phi\phi}} \right) {{\tp}}^{2}
+ \frac32\left( G_{{3,X\phi\phi}}{{\tp}}^{2}- G_{{2,X
\phi}}-2  G_{{3,\phi\phi}}  \right) H {\tp} 
\notag\\
&&\hspace{1cm}
+ \frac32\left( {{\tp}}^{2}G_{{3,X\phi}}+2
  G_{{4,\phi\phi}} \right) {\dot{H}}
  + \frac32\left( 
3  {{\tp}}^{2}G_{{3,X\phi}}+4  G_{{4,\phi\phi}} \right) {H}^{2}
\notag\\
&&\hspace{1cm}
-\left[ \frac12G_{{2,X\phi}}+G_{{3,\phi\phi}} 
-\frac32\left( G_{{3,{XX}  \phi}}{{\tp}}^{2}
+2  G_{{3,X\phi}} \right) H {\tp}+\frac12 \left( 
G_{{2,{XX}  \phi}}+ G_{{3,X\phi\phi}} \right) {{\tp}}^{2}
 \right] {\ttp}\,,
\notag\\
&&
D_4 = - \left( G_{2,X}+2\,G
_{3,\phi} \right) {\tp}
-\left( G_{2,{
XX}}+G_{3,X\phi} \right) {\tp}^{3}
+3\left( 3G_{3,X}{\tp}^{2}+G_{3,{XX}}{\tp}^{4}
+2G_{4,\phi} \right) H\,,\\
& &
B_1=\frac{2}{\tp}\dot{q_t}\,,
\qquad 
B_2=\dot{q_t}+3Hq_t\,,
\qquad 
B_3=\dot{D_6}+3HD_6-D_7\,,
\qquad
B_4=\dot{D_7}+3HD_7\,. 
\label{Bi}
\ea
We recall that the coefficients ${\cal C}_{1,2,3}$ and 
$D_{1,5,6,7}$ are given by 
Eqs.~(\ref{C1})-(\ref{C3}) and 
Eqs.~(\ref{D1})-(\ref{D7}), respectively.
The coefficients $D_{2,3,4}$ can be expressed by 
using $q_t, D_{1,5,6,7}$ and their time derivatives, as
\ba
2\dot{\phi}^2 D_2 &=&
-2H \dot{q}_t -\dot{\phi} 
\left( \dot{D}_6+H D_6+D_7 \right)\,,\label{D2re} \\
2\dot{\phi} D_3 &=&\dot{D}_5+3H D_5-3H \left( 
\dot{D}_7+3H D_7 \right)
-3H\left[ f_{1,\phi}-f_{1,X \phi} \dot{\phi}^2
-\left( f_{1,X}+f_{1,XX} \dot{\phi}^2 \right) \ddot{\phi} 
\right] \rho_c-9H^2 \dot{\phi}f_{1,X}P_c \nonumber \\
& &
+3H n_c \left( 3H n_c f_{2,n_c}-3\dot{\phi} \ddot{\phi}f_{2,X} 
+3H \dot{\phi}^2 n_c f_{2,n_cX} -\dot{\phi}^3 f_{2,X \phi}
-\dot{\phi^3} \ddot{\phi} f_{2,XX}
\right)
\,,\label{D3re}\\
D_4&=&-2\dot{\phi}D_1-3HD_6\,,
\label{D4re}
\ea
where we used Eq.~(\ref{Eqphi}) for the derivation 
of Eq.~(\ref{D3re}). 
The second-order Horndeski action (\ref{SH2}) is written in the gauge-ready form. 
If we choose the flat gauge in which both $\zeta$ and $E$ vanish, we have 
that $L_{\rH}^{\zeta}=0$ and $L_{\rH}^{E}=0$. 
In this case, what is left in ${\cal S}_{\rm H}^{(2)}$ is the Lagrangian $L_{\rm H}^{\rm flat}$ 
alone. As we will see in Sec.~\ref{gaugesec}, we can also choose other gauges 
depending on the problem at hand.

We also expand the sum of actions ${\cal S}_{M}+{\cal S}_{\rm int}$
up to quadratic order in scalar perturbations.
After integrating out the fields $\delj_I$ by using the relation (\ref{deljre}), the second-order 
action in the matter sector is given by the sum of ${\cal S}_{M}^{(2)}$ 
and ${\cal S}_{\rm int}^{(2)}$. The former is expressed in the form 
\be
{\cal S}_M^{(2)}=\int 
{\rm d}t\,{\rm d}^3x \left( L_M^{\rm flat}
+L_M^{\zeta}+L_M^{E} \right)\,,
\ee
where  
\ba
L_M^{\rm flat}
&=&
\sum_{I=c,b,r}
a^3\left[\left\{ (\rho_I+P_I) \frac{\D^2 \chi}{a^2}-\dot\drho
-3H (1+c_I^2) \drho\right\}v_I 
-\frac{1}{2} (\rho_I+P_I) \frac{(\D v_I)^2}{a^2}
-\frac{c_I^2}{2(\rho_I+P_I)} \drho^2 
\right.\notag\\
&&\hspace{1.7cm}
\left.-\alpha \drho -\frac{1}{2}\rho_I\left\{\frac{(\D\chi)^2}{a^2}-\alpha^2\right\}
\right]\,,
\label{LMF}\\
L_M^{\zeta}
&=& -\sum_{I=c,b,r}3a^3\left[
(\rho_I+P_I) v_I\dot{\zeta}
+\rho_I\alpha \zeta-\frac12P_I \zeta^2 
\right]\,,
\label{LMze}\\
L_M^{E}
&=& \sum_{I=c,b,r}a^3\left[
(\rho_I+P_I) (\dot{v_I}-3Hc_I^2v_I)-\rho_I\alpha
+P_I \left(\zeta-\frac12\D^2E\right)
\right]\,\D^2 E\,. 
\label{LME}
\ea
The latter interacting second-order action is given by 
\be
{\cal S}_{\rm int}^{(2)}=\int 
{\rm d}t\,{\rm d}^3x \left( L_{\rm int1}^{\rm flat}
+L_{\rm int1}^{\zeta}+L_{\rm int1}^{E}+
L_{\rm int2}^{\rm flat}
+L_{\rm int2}^{\zeta}+L_{\rm int2}^{E} \right)\,,
\ee
where
\ba
\hspace{-0.6cm}
L_{\rm int1}^{\rm flat}
&=&a^3\left[
\left\{(\rho_c+P_c) \frac{\D^2 \chi}{a^2}-\dot\drhoc
-3H (1+c_c^2) \drhoc\right\} f_1 v_c 
-\frac{1}{2} (\rho_c+P_c) f_1 \frac{(\D v_c)^2}{a^2}
-\frac{c_c^2}{2 (\rho_c+P_c)}f_1 \drhoc^2
\right.\notag\\
&&
\left.
-\left\{(f_1-\tp^2f_{1,X})\alpha+\tp f_{1,X}\dot{\dphi}+f_{1,\phi}\dphi\right\}\drhoc
-\rho_cf_{1,X}\left\{\tp\dphi\frac{\D^2\chi}{a^2}-\frac{(\D\dphi)^2}{2a^2}\right\}
\right.\notag\\
&&
\left.
+\frac12\left\{\rho_c(f_{1,X\phi}+\tp^2f_{1,XX\phi})\ttp
-3HP_c\tp f_{1,X\phi}
-\rho_c(f_{1,\phi\phi}-\tp^2f_{1,X\phi\phi})\right\}\dphi^2
-\rho_c(f_{1,\phi}-\tp^2f_{1,X\phi})\alpha\dphi
\right.\notag\\
&&
\left.
-\frac12\rho_c(f_{1,X}+\tp^2f_{1,XX})(\tp\alpha-\dot{\dphi})^2
-\frac12\rho_c(f_1-\tp^2f_{1,X})\left\{\frac{(\D\chi)^2}{a^2}-\alpha^2\right\}
\right]
\,,
\label{Lint1F}\\
\hspace{-0.6cm}
L_{\rm int1}^{\zeta}
&=& -3a^3\left[
\{(\rho_c+P_c) f_1 v_c-\rho_c\tp f_{1,X}\dphi\}\dot{\zeta}
+\rho_c(f_1-\tp^2f_{1,X})\alpha\zeta
-\frac12P_cf_1\zeta^2\right.\notag\\
&&
\left.+\left\{\rho_c(f_{1,\phi}-\tp^2f_{1,X\phi})
-\rho_c\ttp (f_{1,X}+\tp^2f_{1,XX})
+3HP_c\tp f_{1,X}\right\}\dphi\,\zeta
\right]\,,
\label{Lint1ze}\\
\hspace{-0.6cm}
L_{\rm int1}^{E}
&=& -a^3 \left[ \left\{ (\rho_c+P_c) f_1 v_c-\rho_c\dot{\phi} f_{1,X} \dphi \right\}
\partial^2 \dot{E}
+\rho_c(f_1-\tp^2f_{1,X})\alpha\,\D^2 E
-P_cf_1\zeta\D^2E
+\frac12P_cf_1(\D^2 E)^2\right.\notag\\
&&\left.
+\left\{\rho_c(f_{1,\phi}-\tp^2f_{1,X\phi})
-\rho_c\ttp (f_{1,X}+\tp^2f_{1,XX})
+3HP_c\tp f_{1,X}\right\} \dphi\D^2 E
\right] \,,
\label{Lint1E}
\ea
and
\ba
L_{\rm int2}^{\rm flat}
&=&-a^3n_c\left[
\left\{ (\rho_c+P_c)\frac{\D^2\chi}{a^2}-
\dot{\drhoc}-3H(1+c_c^2)\drhoc \right\} 
\frac{\dot{\phi}n_c f_{2,n_c}}{\rho_c+P_c} v_c
-\dot{\phi} n_cf_{2,n_c}\frac{(\D v_c)^2}{2a^2}
-\frac{\dot{\phi}n_c(2 f_{2,n_c}+n_cf_{2,n_cn_c})}{2(\rho_c+P_c)^2}\drhoc^2
\right.\notag\\
&&\left.
-\frac{1}{\rho_c+P_c}
\left\{ 
n_c f_{2,n_c}\dot{\dphi}-\dot{\phi}^2(f_{2,X}+n_cf_{2,n_c X})(\dot{\phi}\alpha-\dot{\dphi})
+[n_c(3H f_{2,n_c}+\dot{\phi} f_{2,n_c \phi})-f_{2,X}\dot{\phi} \ddot{\phi}] \dphi \right\}\drhoc
\right.\notag\\
&&\left.
-\dot{\phi} f_{2,X} \left\{\dot{\phi} \dphi\frac{\D^2\chi}{a^2}-\frac{(\D \dphi)^2}{2a^2} \right\} 
+\frac12  \left\{\dot{\phi} \ddot{\phi}(3 f_{2,X\phi}+\dot{\phi}^2f_{2,XX\phi})
-3 H n_c(f_{2,n_c \phi}+\dot{\phi}^2f_{2,n_cX\phi}) +\dot{\phi}^3f_{2,X\phi\phi}\right\} \dphi^2
\right.\notag\\
&&\left.
 +\dot{\phi}^3f_{2,X\phi} \alpha\dphi 
-\frac12 \dot{\phi}(3 f_{2,X}+\dot{\phi}^2f_{2,XX})(\dot{\phi}\alpha-\dot{\dphi})^2
+\frac12 \dot{\phi}^3f_{2,X}\left\{\frac{(\D\chi)^{2}}{a^2}-\alpha^2\right\}
\right]
\,,
\label{Lint2F}\\
L_{\rm int2}^{\zeta}
&=& 3a^3n_c\left[
\dot{\phi}(n_c f_{2,n_c} v_c-\dot{\phi} f_{2,X} \dphi) \dot{\zeta}
-\dot{\phi}^3f_{2,X} \alpha \zeta
-\frac12 \dot{\phi} n_c f_{2,n_c} \zeta^2
\right.\notag\\
&&\left.
-\left\{\dot{\phi}\ddot{\phi}(3 f_{2,X}+\dot{\phi}^2f_{2,XX}) 
-3 H n_c (f_{2,n_c}+\dot{\phi}^2f_{2,n_c X})
+\dot{\phi}^3f_{2,X\phi}\right\}\dphi\,\zeta 
\right]\,,
\label{Lint2ze}\\
L_{\rm int2}^{E}
&=& a^3n_c\left[
\dot{\phi}(n_c f_{2,n_c} v_c-\dot{\phi} f_{2,X} \dphi) \D^2\dot{E}
-\dot{\phi} n_c f_{2,n_c}\zeta\,\D^2E
-\dot{\phi}^3f_{2,X} \alpha \D^2E
+\frac12 \dot{\phi} n_c f_{2,n_c} (\D^2E)^2
\right.\notag\\
&&\left.
-\left\{\dot{\phi}\ddot{\phi}(3 f_{2,X}+\dot{\phi}^2f_{2,XX}) 
-3 H n_c (f_{2,n_c}+\dot{\phi}^2f_{2,n_c X})
+\dot{\phi}^3f_{2,X\phi}\right\}\dphi\,\D^2E 
\right]\,.
\label{Lint2E}
\ea

Now, we are ready for computing the total second-order action 
${\cal S}_s^{(2)}={\cal S}_{\rH}^{(2)}+{\cal S}_M^{(2)}+{\cal S}_{\rm int}^{(2)}$. 
On using the background Eqs.~(\ref{Eq00})-(\ref{Eqphi}), the terms 
containing ${\cal C}_1$, ${\cal C}_2$, and ${\cal C}_3$ 
in Eqs.~(\ref{Lgf})-(\ref{LgE}) cancel the sum of
last contributions to Eqs.~(\ref{LMF}), (\ref{Lint1F}), (\ref{Lint2F}), 
last two contributions to Eqs.~(\ref{LMze})-(\ref{LME}), 
and the terms except for $\dot{\zeta}$ and $\D^2\dot{E}$ 
in Eqs.~(\ref{Lint1ze})-(\ref{Lint1E}) and (\ref{Lint2ze})-(\ref{Lint2E}). 
Then, the resulting full second-order action is given by 
\be
{\cal S}_s^{(2)}=\int 
{\rm d}t\,{\rm d}^3x \left( L_{0}+
L_{\rm int} \right)\,,
\label{faction}
\ee
where
\ba
L_0
&=& a^3\biggl\{
D_1\dot{\dphi}^2+D_2\frac{(\partial\dphi)^2}{a^2}+D_3\dphi^2
+\left(D_4\dot{\dphi}+D_5\dphi+D_6\frac{\partial^2\dphi}{a^2}\right) \alpha
-\left(D_6\dot{\dphi}-D_7\dphi\right)\frac{\partial^2\chi}{a^2}
 \nonumber \\
&&
+\left(\dot{\phi} D_6-2Hq_t\right)\alpha\frac{\partial^2\chi}{a^2}
+\left(\dot{\phi}^2 D_1+3H \dot{\phi} D_6-3H^2 q_t\right)\alpha^2
 \nonumber \\
&&
+\sum_{I=c,b,r}\left\{ \left( \rho_I+P_I \right)v_I \frac{\partial^2 \chi}
{a^2}-v_I \dot{\delta \rho}_I-3H (1+c_I^2) v_I \delta \rho_I 
-\frac{\rho_I+P_I }{2a^2} (\partial v_I)^2
-\frac{c_I^2}{2 (\rho_I+P_I)} \delta \rho_I^2 
-\alpha \delta \rho_I \right\} 
\nonumber \\
&&
+\biggl\{
3D_6\dot{\dphi}-3D_7\dphi-3\left(\dot{\phi} D_6-2Hq_t\right)\alpha
-\sum_{I=c,b,r} 3(\rho_I+P_I)v_I
+2q_t\frac{\pa^2\chi}{a^2}\biggr\}\dot{\zeta}-3q_t\dot{\zeta}^2
\nonumber \\
&&
-\left(B_1\dphi+2q_t\alpha\right)\frac{\pa^2\zeta}{a^2}
+q_t \frac{(\pa\zeta)^2}{a^2}
+\bigg[ 2q_t\ddot{\zeta}+2B_2\dot{\zeta}
-D_6\ddot{\dphi}-B_3\dot{\dphi}+B_4\dphi
+\frac{1}{a^3}\frac{{\rm d}}{{\rm d}t}
\left\{a^3 \left(\dot{\phi} D_6-2Hq_t 
\right) \alpha\right\} \nonumber \\
&&
+\sum_{I=c,b,r}(\rho_I+P_I)(\dot{v}_I-3Hc_I^2v_I) \biggr] 
\pa^2 E \biggr\}\,,
\label{L0}
\ea
and 
\ba
L_{\rm int}
&=&a^3\left[
\left(f_1-\frac{\tp n_c^2f_{2,n_c}}{\rho_c+P_c}\right)
\left\{ (\rho_c+P_c) \frac{\D^2 \chi}{a^2}
-\dot\drhoc-3H(1+c_c^2) \drhoc\right\} v_c 
-\left\{(\rho_c+P_c)f_1-\tp n_c^2 f_{2,n_c}\right\} \frac{(\D v_c)^2}{2a^2}
\right.\notag\\
&&
\left. 
-\left\{
\left(f_1-\tp^2f_{1,X}+\frac{\tp^3n_c\{f_{2,X}+n_cf_{2,n_c X}\}}{\rho_c+P_c}\right)\alpha
+\left(\tp f_{1,X}-\frac{n_c\{n_cf_{2,n_c}+\tp^2(f_{2,X}+n_cf_{2,n_c X})\}}{\rho_c+P_c}\right)\dot{\dphi}
\right.\right.\notag\\
&&
\left.\left. 
+\left( f_{1,\phi}-\frac{n_c\{n_c(3Hf_{2,n_c}+\tp f_{2,n_c \phi})
-\tp\ttp f_{2,X}\}}{\rho_c+P_c}\right)\dphi
\right\}\drhoc
-(f_{1,X}\rho_c-\tp n_cf_{2,X})\left\{ \tp\dphi\frac{\D^2\chi}{a^2}-\frac{(\D\dphi)^2}{2a^2}\right\}
\right.\notag\\
&&
\left. 
+\frac12\left\{
\rho_c(f_{1,X\phi}+\tp^2f_{1,XX\phi})\ttp 
-n_c (3f_{2,X\phi}+\tp^2f_{2,XX\phi})\tp \ttp
-\rho_c(f_{1,\phi\phi}-\tp^2f_{1,X\phi\phi})
-3HP_c\tp f_{1,X\phi}
\right.\right.\notag\\
&&
\left.\left. 
+n_c\left(3Hn_c\{f_{2,n_c\phi}+\tp^2f_{2,n_cX\phi}\}
-\tp^3f_{2,X\phi\phi}\right)
\right\} \dphi^2-\frac{1}{2(\rho_c+P_c)}
\left\{f_1c_c^2-\frac{\tp n_c^2(2f_{2,n_c}+n_cf_{2,n_cn_c})}{\rho_c+P_c}\right\}\delta\rho_c^2
\right.\notag\\
&&
\left.
-\left\{\rho_c(f_{1,\phi}-\tp^2f_{1,X\phi})+\tp^3n_cf_{2,X\phi}\right\}\alpha\dphi
-\frac12\left\{\rho_c(f_{1,X}+\tp^2f_{1,XX})-\tp n_c (3f_{2,X}+\tp^2f_{2,XX})\right\}(\tp\alpha-\dot{\dphi})^2
\right.\notag\\
&&
\left.
-\left\{[(\rho_c+P_c) f_1-\tp n_c^2f_{2,n_c}] v_c
-\tp(\rho_c f_{1,X}-\tp n_cf_{2,X}) \delta \phi \right\}
\left( 3 \dot{\zeta}+\partial^2 \dot{E} \right)
\right]\,.
\ea
The Lagrangian $L_{\rm int}$ characterizes the interaction 
between CDM and the scalar field arising from nonvanishing couplings $f_1$ and $f_2$. 

\subsection{Perturbation equations}
\label{scaeqsec}

We derive all the linear perturbation equations of motion 
in Fourier space with the comoving wavenumber $k$.
Varying the action (\ref{faction}) with respect to nondynamical 
fields $\alpha$, $\chi$, $v_I$, and $E$, 
it follows that 
\ba
& & \left[ D_4+\dot{\phi} \left( f_{1,X}+\dot{\phi}^2 f_{1,XX} \right)\rho_c
-n_c\tp^2\left(3f_{2,X}+\tp^2f_{2,XX}\right)\right] 
\dot{\delta \phi}-3\left(\dot{\phi} D_6-2Hq_t\right)\dot{\zeta}
\notag\\
&&
+\left[ D_5-\left( f_{1,\phi}-\dot{\phi}^2 f_{1,X \phi} \right) \rho_c
-\tp^{3} n_cf_{2,X\phi} \right] \delta \phi 
+\left[2\dot{\phi}^2 D_1+6H \dot{\phi} D_6-6H^2 q_t
-\dot{\phi}^2\left( f_{1,X}+\dot{\phi}^2 f_{1,XX} \right) \rho_c
\right.\notag\\
&&\left.
+\tp^3n_c\left(3f_{2,X}+\tp^2f_{2,XX}\right)\right] \alpha
+\frac{k^2}{a^2}\left[ 2q_t\zeta-\left(\dot{\phi} D_6-2Hq_t\right)\left(\chi-a^2\dot{E}\right)
-D_6\dphi \right]-\sum_{I=c,b,r} \delta \rho_I 
\nonumber \\
& &
-\left[ f_1-\dot{\phi}^2 f_{1,X}+\frac{\tp^3n_c(f_{2,X}+n_cf_{2,n_cX})}{\rho_c+P_c} \right] 
\delta \rho_c=0\,,\label{pereq1} \\
& &
D_6\dot{\dphi}-2q_t\dot{\zeta}
-\left( D_7-\dot{\phi}f_{1,X} \rho_c+\tp^2n_cf_{2,X} \right) \dphi-\left(\dot{\phi} D_6-2Hq_t\right)\alpha
-\sum_{I=c,b,r} \left( \rho_I+P_I \right)v_I
\notag\\
&&
-\left[f_1 (\rho_c+P_c)-\tp\,n_c^2\,f_{2,n_c}\right] v_c=0\,,
\label{eqchi}\\
& &
\dot{\delta \rho}_I+3H \left( 1+c_I^2 \right) \delta \rho_I
+3 \left (\rho_I+P_I \right) \dot{\zeta}
+\frac{k^2}{a^2} \left( \rho_I+P_I \right) 
\left( v_I+\chi-a^2\dot{E} \right)=0\,, 
\qquad {\rm for} \quad I=c,b,r\,,\label{pereq3}\\
& &
2q_t\ddot{\zeta}+2B_2\dot{\zeta}
-D_6\ddot{\dphi}-\left( B_3+\dot{\phi} f_{1,X} \rho_c-\tp^2n_cf_{2,X} \right) 
\dot{\dphi}+\left[ B_4-\left\{ \dot{\phi}^2 f_{1,X \phi}
+ \ddot{\phi}\left(f_{1,X}+\dot{\phi}^2 f_{1,XX}\right)\right\} \rho_c
\right.\notag\\
&&\left.
+3HP_c\tp f_{1,X}+\tp\,n_c\left\{\ttp\left(2f_{2,X}+\tp^2f_{2,XX}\right)
+\tp^2f_{2,X\phi}\right\}-3H\tp^2n_c^2f_{2,n_cX}
 \right] \dphi+\frac{1}{a^3}\frac{{\rm d}}{{\rm d}t}
\left\{a^3(\dot{\phi} D_6-2Hq_t)\alpha\right\}
\nonumber \\
& &
+\sum_{I=c,b,r} (\rho_I+P_I)(\dot{v}_I-3Hc_I^2 v_I)
+\left[f_1 (\rho_c+P_c)-\tp\,n_c^2f_{2,n_c}\right]\dot{v}_c
-\left[\left\{3Hc_c^2f_1-\dot{\phi} \left( f_{1,\phi}
+\ddot{\phi} f_{1,X} \right)\right\} (\rho_c+P_c)
\right.\notag\\
&&\left.
+n_c^2\left\{\left(\ttp-3H\tp\right)f_{2,n_c}
+\tp^2\left(f_{2,n_c\phi}+\ttp f_{2,n_cX}\right)\right\}
-3H\tp\,n_c^3f_{2,n_cn_c}
\right] v_c=0\,.
\label{eqE}
\ea
These equations can be used to eliminate $\alpha$, $\chi$, $v_I$, 
and $E$ from the action (\ref{faction}).
In Sec.~\ref{stabsec}, we will derive the stability conditions 
of dynamical perturbations in the small-scale limit after 
the elimination of nondynamical perturbations. 

Variations of the action (\ref{faction}) with respect to dynamical fields 
$\delta \phi$, $\delta \rho_I$, and $\zeta$ lead to
\ba
\hspace{-1cm}
& &
\dot{\cal Z}+3H {\cal Z}+3 \left( D_7-\dot{\phi} f_{1,X} \rho_c 
+\tp^2n_cf_{2,X}\right)\dot{\zeta}+M_{\phi}^2 \delta \phi 
-\left[ \left\{ \ddot{\phi} (f_{1,X\phi}+\dot{\phi}^2 f_{1,XX \phi})
-f_{1,\phi \phi}+\dot{\phi}^2 f_{1,X \phi \phi} \right\} \rho_c
\right.\notag\\
&&\left.
-3HP_c\tp f_{1,X\phi}
-\tp\,n_c\left\{\ttp\left(3f_{2,X\phi}+\tp^2f_{2,XX\phi}\right)
+\tp^2f_{2,X\phi\phi}\right\}
+3Hn_c^2\left(f_{2,n_c\phi}+\tp^2f_{2,n_cX\phi}\right)
\right]\delta \phi 
\nonumber \\
\hspace{-1cm}
& &
+\left[f_{1,\phi}-\frac{n_c}{\rho_c+P_c}\left(3Hn_cf_{2,n_c}
+\tp\,n_cf_{2,n_c\phi}-\tp\ttp f_{2,X}\right)\right]\delta \rho_c
-\left[ D_5-( f_{1,\phi}-\dot{\phi}^2 f_{1,X \phi}) \rho_c 
-\tp^3n_cf_{2,X\phi}\right] \alpha 
\notag\\
 &&
-\frac{k^2}{a^2} \left[ 2D_2 \delta \phi -D_6 \alpha 
-D_7 \chi+B_1\zeta-a^2B_4E
+\left(f_{1,X}\rho_c-\tp\,n_cf_{2,X}\right)  
\left\{ \delta \phi+\dot{\phi} (\chi-a^2\dot{E})  
\right\} \right]=0\,,\label{calZeq}\\
\hspace{-1cm}
&&
\left(1+f_1-\frac{\tp\,n_c^2 f_{2,n_c}}{\rho_c+P_c} \right) \dot{v}_c
-\left[3Hc_c^2(1+f_1)-\dot{\phi} \left( f_{1,\phi}+\ddot{\phi}f_{1,X} \right)
+\frac{n_c^2}{\rho_c+P_c}\left\{\ttp f_{2,n_c}-3H\tp(f_{2,n_c}+n_cf_{2,n_cn_c})
\right.\right.\notag\\
&&\left.\left.
+\tp^2\left(f_{2,n_c\phi}+\ttp f_{2,n_cX}\right)\right\}\right]v_c
-\left[ 1+f_1-\dot{\phi}^2f_{1,X}+\frac{\tp^3n_c(f_{2,X}
+n_cf_{2,n_cX})}{\rho_c+P_c} \right] \alpha
\notag\\
&&
-\left[\dot{\phi} f_{1,X}-\frac{n_c(n_cf_{2,n_c}
+\tp^2f_{2,X}+\tp^2\,n_cf_{2,n_cX})}{\rho_c+P_c}\right] \dot{\delta \phi}
-\left[f_{1,\phi}-\frac{n_c(3Hn_cf_{2,n_c}+\tp\,n_c f_{2,n_c\phi}
-\tp\ttp f_{2,X})}{\rho_c+P_c}\right] \delta \phi
\notag\\
&&
-\frac{1}{\rho_c+P_c}\left[(1+f_1)c_c^2-\frac{\tp\,n_c^2(2f_{2,n_c}+n_cf_{2,n_cn_c})}{\rho_c+P_c}\right] \drhoc
=0\,,
\label{vceq}\\
\hspace{-1cm}
& &
\dot{v}_I-3H c_I^2\,v_I-\frac{c_I^2}{\rho_I+P_I} \delta \rho_I 
-\alpha=0\,,\qquad {\rm for} \quad I=b,r\,,
\label{veq}\\
\hspace{-1cm}
& &
\dot{\cal W} +3H{\cal W}+\sum_{I=c,b,r} (\rho_I+P_I)
(\dot{v}_I-3Hc_I^2\,v_I)
+ \left[ (\rho_c+P_c)f_1-\tp\,n_c^2f_{2,n_c}\right] \dot{v}_c
+\frac{k^2}{3a^2}\left(2q_t\alpha+2q_t\zeta
+B_1\dphi\right) \notag\\
&&
-\left[n_c^2\left\{\ttp f_{2,n_c}-3H\tp(f_{2,n_c}+n_cf_{2,n_cn_c})
+\tp^2(f_{2,n_c\phi}+\ttp f_{2,n_cX})\right\} \right. \notag\\
&&
\left.
+(\rho_c+P_c)\left\{3Hc_c^2f_1-\dot{\phi} (f_{1,\phi}
+\ddot{\phi}f_{1,X})\right\} 
\right] v_c=0\,,
\label{calWeq}
\ea
where 
\ba
M_{\phi}^2 &\equiv& -2D_3\,,\\
{\cal Z} &\equiv& \left[ 2D_1-(f_{1,X}+\dot{\phi}^2 f_{1,XX}) \rho_c
+\tp n_c (3f_{2,X}+\tp^2 f_{2,XX}) 
\right] \dot{\delta \phi}+3D_6\dot{\zeta}
+\left[ D_4 +\dot{\phi} (f_{1,X}+\dot{\phi}^2 f_{1,XX})\rho_c
\right.\notag\\
&&\left.
-\tp^2 n_c (3f_{2,X}+\tp^2f_{2,XX}) \right]\alpha 
-\left[\dot{\phi} f_{1,X}-\frac{n_c(n_c f_{2,n_c}
+\tp^2f_{2,X}+\tp^2n_cf_{2,n_cX})}{\rho_c+P_c}\right] \delta \rho_c 
\nonumber \\
& &
+\frac{k^2}{a^2}\left[D_6\chi 
-a^2(D_6\dot{E}+D_7E)\right]\,, 
\label{Zdef}\\
{\cal W} &\equiv& 2q_t\dot{\zeta}-D_6\dot{\dphi}
+\left( D_7-\dot{\phi} f_{1,X} \rho_c+\tp^2 n_c f_{2,X} \right)\dphi
+\left(\dot{\phi} D_6-2Hq_t\right)\alpha
+\frac{2k^2}{3a^2} q_t (\chi-a^2 \dot{E})\,. 
\label{Wdef}
\ea
The coefficient $-2D_3$ contains the term $-G_{2,\phi \phi}$. 
For a canonical scalar with the potential $V(\phi)$, i.e., $G_2=X-V(\phi)$,
the term $-G_{2,\phi \phi}$ reduces to $V_{,\phi \phi}$, 
which corresponds to the field mass squared. 
In Ref.~\cite{Boehmer:2015sha}, the scalar perturbation equations were 
derived in coupled quintessence with the interacting Lagrangian 
${\cal L}_{\rm int}=q J^{\mu}_c\partial_{\mu}\phi/n_c$, 
where $q$ is a constant. 
They can be recovered by setting $f_1=0$, $f_2=q/n_c$, 
$G_2=X-V$, $G_3=0$, $G_4=M_{\rm pl}^2/2$, $\rho_b=0$, and 
$\rho_r=0$ in Eqs.~(\ref{pereq1})-(\ref{pereq3}), (\ref{calZeq}), 
(\ref{vceq}), and (\ref{calWeq}). 

On using Eqs.~(\ref{eqE}) and (\ref{calWeq}), the time derivatives 
$\ddot{\zeta}$ and $\ddot{\dphi}$ are eliminated to give
\be
q_t \left[ \alpha+\dot{\chi}
+\zeta+H\chi-a^2 (\ddot{E}+3 H \dot{E})  
\right]+  \dot{q}_t \left( \chi-a^2 \dot{E} 
+\frac{\dphi}{\dot{\phi}} \right)=0\,,
\label{eqE2}
\ee
where we used the relation $B_1=2\dot{q}_t/\dot{\phi}$.
The above perturbation equations of motion can be applied to 
any choices of gauges.

\section{Small-scale stability conditions}
\label{stasec}

By using the second-order action of scalar perturbations derived in 
Sec.~\ref{actionsec}, we identify conditions for the absence of ghosts 
and Laplacian instabilities in the small-scale limit. 
Before doing so, we introduce commonly used gauge-invariant variables 
and discuss several different choices of gauges.

\subsection{Gauge-invariant variables and gauge fixings}
\label{gaugesec}

Let us consider the infinitesimal gauge transformation of time 
and spatial coordinates of the forms
$\tilde{t}=t+\xi^{0}$ and $\tilde{x}^{i}=x^{i}+\delta^{ij} \partial_{j} \xi$, 
where $\xi^{0}$ and $\xi$ are scalar quantities. 
Then, the four scalar metric perturbations in the line element 
(\ref{permet}) transform as \cite{Bardeen,Kodama:1985bj,Mukhanov:1990me}
\be
\tilde{\alpha}=\alpha-\dot{\xi}^{0}\,,\qquad 
\tilde{\chi}=\chi+\xi^{0}-a^2 \dot{\xi}\,,\qquad 
\tilde{\zeta}=\zeta-H \xi^{0}\,,\qquad 
\tilde{E}=E-\xi\,.
\label{gaugetra1}
\ee
The transformations of $\delta \phi$, $\delta \rho_I$, and $v_I$ are 
given, respectively, by 
\be
\widetilde{\delta \phi}=\delta \phi-\dot{\phi}\,\xi^{0}\,,
\qquad 
\widetilde{\delta \rho_I}=\delta \rho_I-\dot{\rho}_I \xi^{0}\,,
\qquad 
\tilde{v}_I=v_I-\xi^{0}\,.
\label{gaugetra2}
\ee
Then, the following perturbed quantities are invariant under the 
gauge transformation, 
\ba
& &
\delta \phi_{\rm f}=\delta \phi
-\frac{\dot{\phi}}{H}\zeta\,,\qquad 
\delta \rho_{I{\rm f}}=\delta \rho_I -\frac{\dot{\rho}_I}{H}\zeta\,,
\qquad 
v_{I{\rm f}}=v_I-\frac{\zeta}{H}\,,
\\
& & 
{\cal R}=\zeta-\frac{H}{\dot{\phi}} \delta \phi\,,\qquad 
\delta \rho_{I{\rm u}}=\delta \rho_I -\frac{\dot{\rho}_I}
{\dot{\phi}}\delta \phi\,,\qquad 
v_{I{\rm u}}=v_I-\frac{\delta \rho_I}{\dot{\rho}_I}\,,
\\
& & 
\delta \phi_{\rm N}=\delta \phi+\dot{\phi}\left(\chi-a^2 \dot{E}\right)\,,
\qquad 
\delta \rho_{I\rm N}=\delta \rho_I+\dot{\rho}_I \left(\chi-a^2 \dot{E}\right)\,,
\qquad 
v_{I{\rm N}}=v_I+\chi-a^2 \dot{E}\,.
\label{delphiN}
\ea
The gauge-invariant gravitational potentials first introduced 
by Bardeen \cite{Bardeen} are given by 
\be
\Psi=\alpha+\frac{{\rm d}}{{\rm d}t} 
\left( \chi - a^2 \dot{E} \right)\,,\qquad 
\Phi=\zeta+H \left( \chi - a^2 \dot{E} \right)\,.
\label{PsiPhi}
\ee
We recall that the background CDM density containing the effect 
of interactions is given by $\hat{\rho}_c=(1+f_1)\rho_c$. 
We also introduce the corresponding gauge-invariant CDM 
density perturbation, as 
\be
\widehat{\delta \rho}_{c{\rm N}}=
\left( 1+f_1 \right) \delta \rho_{c{\rm N}}
+\left[ f_{1,\phi} \delta \phi_{\rm N}
+f_{1,X} \dot{\phi} \left( \dot{\delta \phi}_{\rm N} 
-\dot{\phi} \Psi \right) \right] \rho_c\,.
\label{delrhohat}
\ee

The gauge choice corresponds to fixing the infinitesimal
scalars $\xi^0$ and $\xi$. 
The latter can be fixed by choosing 
\be
E=0\,.
\ee
There are several different gauge choices for the fixing 
of $\xi^0$. The representative examples are 
\ba
& &
{\rm (i)}~\zeta=0 \qquad \quad ({\rm Flat~gauge})\,,
\label{flat}\\
& &
{\rm (ii)}~\delta \phi=0 \qquad\,({\rm Unitary~gauge})\,,
\label{unitary}\\
& &
{\rm (iii)}~\chi=0 \qquad~({\rm Newtonian~gauge})\,.
\ea
In the flat gauge, the dynamical scalar perturbations are 
given by $\delta \phi_{\rm f}=\delta \phi$ and 
$\delta \rho_{I{\rm f}}=\delta \rho_I$, while, in the 
unitary gauge, they correspond to ${\cal R}=\zeta$ and 
$\delta \rho_{I{\rm u}}=\delta \rho_I$. 
We note that the comoving curvature perturbation ${\cal R}$ \cite{Lukash,Lyth} 
is related to the Mukhanov-Sasaki variable $\delta \phi_{\rm f}$ \cite{Muk85,Sasaki},
as ${\cal R}=-(H/\dot{\phi}) \delta \phi_{\rm f}$. 
In the Newtonian gauge, the perturbations in Eqs.~(\ref{delphiN}) and 
(\ref{PsiPhi}) reduce, respectively, to 
$\delta \phi_{\rm N}=\delta \phi$, $\delta \rho_{I\rm N}=\delta \rho_I$, 
$v_{I{\rm N}}=v_I$, $\Psi=\alpha$, and $\Phi=\zeta$. 
For this gauge choice, the gauge-invariant 
dynamical scalar perturbations are ${\cal R}=\Phi-(H/\dot{\phi}) 
\delta \phi_{\rm N}$ (or $\delta \phi_{\rm f}=\delta \phi_{\rm N}
-(\dot{\phi}/H) \Phi$) 
and $\delta \rho_{I\rm N}$.

\subsection{Stability conditions in the small-scale limit}
\label{stabsec}

In order to obtain conditions for the absence of ghosts and Laplacian instabilities, 
we choose two different gauges and show that the small-scale 
stability conditions are independent of the choice of gauges.

\subsubsection{Flat gauge}

Let us begin with the flat gauge characterized by $\zeta=0$ and $E=0$.
We first solve Eqs.~(\ref{pereq1})-(\ref{pereq3}) for nondynamical perturbations 
$\alpha$, $\chi$, $v_c$, $v_b$, $v_r$ and substitute them into 
Eq.~(\ref{faction}). 
After the integration by parts, the resulting second-order action 
is expressed in the form 
\be
{\cal S}_s^{(2)}=\int {\rm d}t\,{\rm d}^3x\,a^{3}\left( 
\dot{\vec{\mathcal{X}}}^{t}{\bm K}\dot{\vec{\mathcal{X}}}
-\frac{k^2}{a^2}\vec{\mathcal{X}}^{t}{\bm G}\vec{\mathcal{X}}
-\vec{\mathcal{X}}^{t}{\bm M}\vec{\mathcal{X}}
-\frac{k}{a}\vec{\mathcal{X}}^{t}{\bm B}\dot{\vec{\mathcal{X}}}
\right)\,,
\label{Ss2}
\ee
where ${\bm K}$, ${\bm G}$, ${\bm M}$, ${\bm B}$ 
are $4 \times 4$ matrices, and 
\be
\vec{\mathcal{X}}^{t}=\left(\dphi_{\rm f}, 
\delta \rho_{c{\rm f}}/k, 
\delta \rho_{b{\rm f}}/k, 
\delta \rho_{r{\rm f}}/k \right) \,.
\label{calX}
\ee
The leading-order terms in the components of matrix
${\bm M}$ are of order $k^0$. 
Taking the small-scale limit ($k \to \infty$), the nonvanishing matrix 
components of ${\bm K}$ and ${\bm G}$ are given by 
\ba
\hspace{-0.8cm}
&&
K_{11}^{(\rm f)}=\frac{H^2 q_t q_s}{(2Hq_t-\tp D_6)^2}\,,
\qquad 
K_{22}^{(\rm f)}=\frac{a^2}{2 (\rho_c+P_c)}
\left(1+f_1-\frac{\tp\,n_c^2f_{2,n_c}}{\rho_c+P_c}\right)\,,
\notag\\
\hspace{-0.8cm}
&& 
K_{33}^{(\rm f)}=\frac{a^2}{2 (\rho_b+P_b)}\,,
\qquad 
K_{44}^{(\rm f)}=\frac{a^2}{2 (\rho_r+P_r)}\,,\label{KGf0}\\
\hspace{-0.8cm}
&&
G_{11}^{(\rm f)}=-D_2+\frac{D_6 D_7-[\sum_{I=c,b,r} (\rho_I+P_I)
+f_1(\rho_c+P_c)-\tp\,n_c^2 f_{2,n_c} ]{\cal G}_1}{2H q_t-\dot{\phi}D_6} 
+\dot{{\cal G}}_1+H{\cal G}_1 \nonumber \\
\hspace{-0.8cm}
& &
\qquad \quad\,
-\frac{(f_{1,X} \rho_c-\tp\,n_c f_{2,X})(2H q_t+\dot{\phi}D_6)}{2(2H q_t-\dot{\phi}D_6)},\nonumber \\
\hspace{-0.8cm}
& &
G_{22}^{(\rm f)}=\frac{a^2}{2(\rho_c+P_c)}\left[(1+f_1)c_c^2
-\frac{\tp\,n_c^2(2f_{2,n_c}+n_cf_{2,n_cn_c})}{\rho_c+P_c}\right]\,,\qquad 
G_{33}^{(\rm f)}=\frac{a^2c_b^2}{2(\rho_b+P_b)}\,,\qquad
G_{44}^{(\rm f)}=\frac{a^2c_r^2}{2(\rho_r+P_r)}\,,
\label{KGf}
\ea
where  
\be
{\cal G}_1=\frac{D_6^2}{2(2Hq_t-\tp D_6)}\,,
\ee
and we used the relation (\ref{D4re}).
The leading-order contributions to the anti-symmetric matrix ${\bm B}$ 
are\footnote{After this paper was published in Phys.\ Rev.\ D {\bf 101}, 063511 (2020), 
we noticed that the propagation speed squared of the scalar field $c_s^2$ is 
affected by the off-diagonal components of ${\bm B}$. The value of $\hat{c}_s^2$ 
defined in Eq.~(\ref{css}) is identical to $c_s^2$ used in the published version. 
As we explicitly show in this arXiv version, 
the off-diagonal component $B_{12}^{{\rm (f)}}$ 
leads to the modification $\Delta c_s^2=(B_{12}^{(\rm f)})^2/
(K_{11}^{(\rm f)}K_{22}^{(\rm f)})$ to $\hat{c}_s^2$, 
where $\Delta c_s^2$ is positive as long as the no-ghost conditions 
(\ref{qscon}) and (\ref{CDMcon}) are satisfied.
} 
\be
B_{12}^{{\rm (f)}}=-B_{21}^{{\rm (f)}}=
-\frac{aH q_t [\dot{\phi} f_{1,X} (\rho_c+P_c)
-\dot{\phi}^2 n_c (f_{2,X}+n_c f_{2,n_c X})-n_c^2 f_{2,n_c}]}
{(2H q_t-\dot{\phi} D_6)(\rho_c+P_c)}\,,
\label{B12}
\ee
while the other components of ${\bm B}$ are lower than the order $k^0$.

We recall that $q_t$ and $q_s$ are defined, respectively, by 
Eqs.~(\ref{qt}) and (\ref{qs}).  
The scalar ghost associated with the field perturbation $\dphi_{\rm f}$ 
is absent for $K_{11}^{(\rm f)}>0$. 
Provided that there is no ghost in the tensor sector ($q_t>0$), 
the condition $K_{11}^{(\rm f)}>0$ translates to 
\be
q_s>0\,.
\label{qscon}
\ee
Since $q_s$ contains the $X$ derivatives of $f_1$ and $f_2$, 
the $X$ dependence in $f_1$ and $f_2$ affects the no-ghost 
condition of $\dphi_{\rm f}$.

For the CDM perturbation $\delta \rho_{c{\rm f}}$, the ghost 
is absent if $K_{22}^{(\rm f)}>0$, i.e., 
\be
q_c\equiv1+f_1-\frac{\tp\,n_c^2f_{2,n_c}}{\rho_c+P_c}>0\,.
\label{CDMcon}
\ee
Hence the coupling $f_1$ and the $n_c$ dependence in $f_2$ lead to 
the modification to $q_c$.
For baryons and radiations, the positivities of $K_{33}^{(\rm f)}$ and 
$K_{44}^{(\rm f)}$ are ensured under the weak energy 
conditions (\ref{weak}).

Let us proceed to the discussion of the propagation speeds 
of perturbations (\ref{calX}), 
which are associated with conditions for 
the absence of Laplacian instabilities.
We need to caution that the nonvanishing components 
$B_{12}^{(\rm f)}$ and $B_{21}^{(\rm f)}$ 
in Eq.~(\ref{B12}) affect the propagation 
speeds of the perturbations 
${\cal X}_1 \equiv \delta \phi_{\rm f}$ and 
${\cal X}_2 \equiv \delta \rho_{c{\rm f}}/k$. 
In the small-scale limit, the equations of motion for ${\cal X}_1$ 
and ${\cal X}_2$ following from the Lagrangian (\ref{Ss2}) 
are of the forms,
\ba
& &
K_{11}^{(\rm f)} \ddot{\cal X}_1+\frac{k^2}{a^2} G_{11}^{(\rm f)} {\cal X}_1
+\frac{k}{a} B_{12}^{(\rm f)} \dot{\cal X}_2 \simeq 0\,,
\label{Eqsma1} \\
& &
K_{22}^{(\rm f)} \ddot{\cal X}_2+\frac{k^2}{a^2} G_{22}^{(\rm f)} {\cal X}_2
+\frac{k}{a} B_{21}^{(\rm f)} \dot{\cal X}_1 \simeq 0\,,
\label{Eqsma2}
\ea
where we neglected the terms lower than the order of $\omega^2$, 
$\omega k$, and $k^2$ in the dispersion relation ($\omega$ is a frequency). 
Substituting the solutions ${\cal X}_i=\tilde{{\cal X}}_i e^{i (\omega t-kx)}$
with $i=1, 2$ into Eqs.~(\ref{Eqsma1})-(\ref{Eqsma2}), where $\tilde{{\cal X}}_i$ 
are constants, it follows that 
\ba
& & \omega^2 \tilde{{\cal X}}_1-\hat{c}_s^2\frac{k^2}{a^2}
\tilde{{\cal X}}_1-i \omega \frac{k}{a} \frac{B_{12}^{(\rm f)}}{K_{11}^{(\rm f)}} 
\tilde{{\cal X}}_2 =0\,,\label{dis1}\\
& & \omega^2 \tilde{{\cal X}}_2-\hat{c}_c^2\frac{k^2}{a^2}
\tilde{{\cal X}}_2-i \omega \frac{k}{a} \frac{B_{21}^{(\rm f)}}{K_{22}^{(\rm f)}} 
\tilde{{\cal X}}_1 =0\,,\label{dis2}
\ea
where 
\ba
\hat{c}_s^2 &=&
\frac{G_{11}^{(\rm f)}}{K_{11}^{(\rm f)}}\,,\label{css}\\ 
\hat{c}_c^2 &=&
\frac{G_{22}^{(\rm f)}}{K_{22}^{(\rm f)}}
=c_c^2-\frac{[(2-c_c^2)f_{2,n_c}+n_c f_{2,n_c n_c}]
\dot{\phi}\,n_c^2}
{(1+f_1)(\rho_c+P_c)-\dot{\phi}\,
n_c^2 f_{2,n_c}}\,.
\label{csc}
\ea
If the coupling $f$ is absent, we have $B_{12}^{(\rm f)}=-B_{21}^{(\rm f)}=0$ 
and hence the propagation speeds of ${\cal X}_1$ and 
${\cal X}_2$ reduce to those given in Eqs.~(\ref{css}) and (\ref{csc}), respectively. 
The interaction between these two fields gives rise to the mixing 
terms in Eqs.~(\ref{dis1}) and (\ref{dis2}). 
In the following, we consider the interacting theories 
obeying the condition
\be
\hat{c}_c^2=0\,.
\label{cccon}
\ee
For CDM satisfying $c_c^2=0$, Eq.~(\ref{csc}) shows that 
the coupling $f_2$ obeys the relation 
$2f_{2,n_c}+n_cf_{2,n_c n_c}=0$, i.e., 
\be
f_2 \propto n_c^{-1}\,.
\label{f2con}
\ee
Substituting Eq.~(\ref{cccon}) into Eq.~(\ref{dis2}), it follows that 
\ba
& &
\omega=0\,,\label{branch1}\\
& &
\omega \tilde{\cal X}_2-i \frac{k}{a}
\frac{B_{21}^{(\rm f)}}{K_{22}^{(\rm f)}} 
\tilde{{\cal X}}_1=0\,.\label{branch2}
\ea
The solution (\ref{branch1}) corresponds to the dispersion relation 
for CDM perturbations, so that the CDM propagation speed squared $c_{\rm CDM}^2$ 
is identical to $0$. 
Then, there is no additional pressure affecting the 
growth of CDM density perturbations. 
Substituting the other solution (\ref{branch2}) into 
Eq.~(\ref{dis1}), we obtain the dispersion relation $\omega^2=c_s^2 k^2/a^2$ 
for the scalar perturbation $\delta \phi_{\rm f}$, with 
\be
c_s^2=\hat{c}_s^2+\Delta c_s^2\,,
\label{cstotal}
\ee
where $\hat{c}_s^2$ is given by Eq.~(\ref{css}), and 
\be
\Delta c_s^2=\frac{(B_{12}^{(\rm f)})^2}
{K_{11}^{(\rm f)}K_{22}^{(\rm f)}}\,.
\label{delcs}
\ee
The Laplacian instability is absent for
\be
c_s^2 \geq 0\,.
\label{cscon}
\ee
Under the no-ghost conditions $K_{11}^{(\rm f)}>0$ 
and $K_{22}^{(\rm f)}>0$, the correction term $\Delta c_s^2$ 
arising from the off-diagonal components of ${\bm B}$ 
is always positive. This means that, as long as $\hat{c}_s^2 \geq 0$, 
the condition (\ref{cscon}) always holds.

The propagation speeds of baryon and radiation density perturbations 
are not affected by the off-diagonal components of ${\bm B}$, so they are given, 
respectively, by $c_b^2=G_{33}^{(\rm f)}/K_{33}^{(\rm f)}$
and $c_r^2=G_{44}^{(\rm f)}/K_{44}^{(\rm f)}$, both of which 
should be non-negative to avoid Laplacian instabilities.

\subsubsection{Unitary gauge}

We also derive stability conditions of scalar perturbations 
by choosing the unitary gauge characterized by $\delta \phi=0$ 
and $E=0$. After eliminating the nondynamical perturbations 
$\alpha$, $\chi$, $v_c$, $v_b$, $v_r$ on account of 
Eqs.~(\ref{pereq1})-(\ref{pereq3}), the second-order scalar 
action (\ref{faction}) reduces to the form (\ref{Ss2}), 
with the dynamical perturbations given by 
\be
\vec{\mathcal{X}}^{t}=\left( 
{\cal R},  
\delta \rho_{c{\rm u}}/k, 
\delta \rho_{b{\rm u}}/k, 
\delta \rho_{r{\rm u}}/k \right) \,.
\label{calX2}
\ee
In the small-scale limit, the nonvanishing components of ${\bm K}$ and 
${\bm G}$ are again diagonal terms, which are given by 
\ba
& &
K_{11}^{(\rm u)}=\frac{\tp^2 q_t q_s}{(2Hq_t-\tp D_6)^2}\,,\qquad
K_{22}^{(\rm u)}=K_{22}^{(\rm f)}\,,\qquad 
K_{33}^{(\rm u)}=K_{33}^{(\rm f)}\,,\qquad 
K_{44}^{(\rm u)}=K_{44}^{(\rm f)}\,,\\
& &
G_{11}^{(\rm u)}=-q_t-\frac{[\sum_{I=c,b,r}(\rho_I+P_I)
+f_1(\rho_c+P_c)-\tp\,n_c^2 f_{2,n_c}]
{\cal G}_2}{2H q_t-\dot{\phi}D_6}
+\dot{\cal G}_2+H{\cal G}_2\,, \nonumber \\
& &
G_{22}^{(\rm u)}=G_{22}^{(\rm f)}\,,\qquad 
G_{33}^{(\rm u)}=G_{33}^{(\rm f)}\,,\qquad 
G_{44}^{(\rm u)}=G_{44}^{(\rm f)}\,,
\label{G11u}
\ea
where 
\be
{\cal G}_2=\frac{2q_t^2}{2Hq_t-\tp D_6}\,.
\ee
The leading-order terms to the matrix components of ${\bm B}$ are 
\be
B_{12}^{{\rm (u)}}=-B_{21}^{{\rm (u)}}=
\frac{a \dot{\phi} q_t [\dot{\phi} f_{1,X} (\rho_c+P_c)
-\dot{\phi}^2 n_c (f_{2,X}+n_c f_{2,n_c X})-n_c^2 f_{2,n_c}]}
{(2H q_t-\dot{\phi} D_6)(\rho_c+P_c)}\,.
\label{B12d}
\ee
The stability conditions of baryon and radiation density perturbations 
are the same as those derived in the flat gauge.
For the perturbations ${\cal R}$ and $\delta \rho_{c{\rm u}}/k$, 
it is convenient to notice the following relations,
\be
\frac{K_{11}^{(\rm u)}}{K_{11}^{(\rm f)}}
=\frac{\dot{\phi}^2}{H^2}\,,\qquad 
\frac{K_{22}^{(\rm u)}}{K_{22}^{(\rm f)}}
=1\,,\qquad 
\frac{G_{11}^{(\rm u)}}{G_{11}^{(\rm f)}}
=\frac{\dot{\phi}^2}{H^2}\,,\qquad 
\frac{G_{22}^{(\rm u)}}{G_{22}^{(\rm f)}}
=1\,,\qquad 
\frac{B_{12}^{(\rm u)}}{B_{12}^{(\rm f)}}
=-\frac{\dot{\phi}}{H}\,,
\label{KG}
\ee
where we used Eqs.~(\ref{Eq00}), (\ref{Eq11}), and (\ref{D2re}) 
for the derivation of the third equality. 
This means that the no-ghost conditions $K_{11}^{(\rm u)}>0$ and 
$K_{22}^{(\rm u)}>0$ in the unitary gauge are identical to those 
obtained in the flat gauge. 

For the interacting theories satisfying the condition 
$\hat{c}_c^2=G_{22}^{(\rm u)}/K_{22}^{(\rm u)}
=G_{22}^{(\rm f)}/K_{22}^{(\rm f)}=0$, 
we obtain the same dispersion relation of CDM as Eq.~(\ref{branch1}), 
so that $c_{\rm CDM}^2$ vanishes.
The propagation speed squared for ${\cal R}$ is given by Eq.~(\ref{cstotal}), 
with
\be
\hat{c}_s^2=\frac{G_{11}^{(\rm u)}}{K_{11}^{(\rm u)}}
=\frac{G_{11}^{(\rm f)}}{K_{11}^{(\rm f)}}\,,\qquad 
\Delta c_s^2=\frac{(B_{12}^{(\rm u)})^2}
{K_{11}^{(\rm u)}K_{22}^{(\rm u)}}
=\frac{(B_{12}^{(\rm f)})^2}
{K_{11}^{(\rm f)}K_{22}^{(\rm f)}}\,,
\ee
where we used Eq.~(\ref{KG}).
Thus, $c_s^2$ is equivalent to each other 
in both unitary and flat gauges.

The matrix component $G_{11}^{(\rm u)}$ in Eq.~(\ref{G11u})
contains the time derivative of ${\cal G}_2$, which generates
the terms $\dot{q}_t$, $\dot{D}_6$, and $\ddot{\phi}$ 
in $\hat{c}_s^2$. We eliminate these time derivatives by using 
Eqs.~(\ref{Eq00}), (\ref{Eq11}), (\ref{D2re}), and the relation 
$B_1=2\dot{q}_t/\dot{\phi}$.
Moreover, it is possible to express $\Delta c_s^2$ in a compact form 
by using $q_s$ and $q_c$.
Then, the propagation speed squared of the scalar field 
is given by $c_s^2=\hat{c}_s^2+\Delta c_s^2$, where
\ba
\hat{c}_s^2
&=& -\frac{D_6^2+2B_1D_6+2q_t 
(2D_2+ f_{1,X}\rho_c-\tp\,n_cf_{2,X})}{q_s}\,,
\label{csfinal}\\
\Delta c_s^2
&=& \frac{2q_t[\dot{\phi} f_{1,X} (\rho_c+P_c)
-\dot{\phi}^2 n_c (f_{2,X}+n_c f_{2,n_c X})-n_c^2 f_{2,n_c}]^2}
{q_s q_c (\rho_c+P_c)}\,.
\label{csfinalde}
\ea
The couplings $f_1$ and $f_2$ nontrivially modify the value of $c_s^2$ 
in comparison to uncoupled theories.

The above discussion shows that, in the small-scale limit, 
the conditions for the absence of ghosts and Laplacian instabilities 
are independent of the choice of gauges. 
Indeed, the analysis in the Newtonian 
gauge also leads to the same stability conditions as 
those derived above (as performed for the uncoupled case 
in Ref.~\cite{Kase:2018aps}).

\section{Effective gravitational couplings for nonrelativistic matter}
\label{Geffsec}

We proceed to the derivation of  the effective gravitational couplings felt by 
CDM and baryons for perturbations deep inside the sound horizon. 
Since we are interested in the cosmological dynamics in the late 
Universe, we ignore the contribution of radiations to the background and 
perturbation equations of motion. 
For CDM and baryons, we consider nonrelativistic matter satisfying
\be
P_I=0\,,\qquad c_I^2=0\,,\qquad 
{\rm for}~~I=c,b\,.
\label{PI}
\ee
Even under the conditions (\ref{PI}), the $n_c$ dependence 
in $f_2$ gives rise to the nonvanishing value of $\hat{c}_c^2$ 
for CDM while the effective pressure $\hat{P}_c$ vanishes. 
We are primarily interested in interacting theories with $\hat{c}_c^2=0$, 
but we will keep $\hat{c}_c^2$ for the derivation of effective gravitational couplings.
{}From Eq.~(\ref{delphiN}), we define the gauge-invariant 
matter density contrast, 
\be
\delta_{I{\rm N}}\equiv \frac{\delta \rho_{I\rm N}}{\rho_I}
=\frac{\delta \rho_I}{\rho_I}
-3H \left( \chi-a^2 \dot{E} \right)\,,
\qquad 
{\rm for}~~I=c,b\,,
\label{delI}
\ee
where we used the continuity Eq.~(\ref{coneq}).
{}From Eq.~(\ref{pereq3}), the matter density contrast obeys
\be
\dot{\delta}_{I{\rm N}}
+\frac{k^2}{a^2} v_{I{\rm N}}+3\dot{\Phi}=0\,,
\label{delIeq}
\ee
where $v_{I{\rm N}}$ and $\Phi$ are gauge-invariant quantities 
defined in Eqs.~(\ref{delphiN}) and (\ref{PsiPhi}).

{}From Eqs.~(\ref{vceq}) and (\ref{veq}), the velocity potentials 
of CDM and baryons satisfy
\ba
&&
\dot{v}_{c{\rm N}}+a_1Hv_{c{\rm N}}
-(1+a_2)\Psi+\frac{H}{\tp}\left[a_2\frac{\dot{\dphi}_{\rm N}}{H}
-(3\hat{c}_c^2+a_1+a_2 \epsilon_{\phi})\dphi_{\rm N}\right]
-\hat{c}_c^2\delta_{c{\rm N}}=0\,,
\label{vceq2}\\
&& 
\dot{v}_{b{\rm N}}-\Psi=0\,, 
\label{vbeq2}
\ea
respectively, where 
\be
a_1 = \frac{\dot{q}_c}{Hq_c}\,,\qquad
a_2 = -\frac{\dot{\phi}[\rho_c \dot{\phi} f_{1,X} 
-\dot{\phi}^2 n_c (f_{2,X}+n_c f_{2,n_c X})-n_c^2 f_{2,n_c}]}
{(1+f_1) \rho_c-\dot{\phi}\,n_c^2 f_{2,n_c}}\,,\qquad
\epsilon_{\phi} = \frac{\ddot{\phi}}{H \dot{\phi}}\,.
\ee
Taking the time derivatives of Eq.~(\ref{delIeq}) for $I=c,b$ 
and using Eqs.~(\ref{vceq2}) and (\ref{vbeq2}), 
it follows that 
\ba
& &
\ddot{\delta}_{c{\rm N}}+\left( 2+a_1 \right)H \dot{\delta}_{c{\rm N}}
+\hat{c}_c^2 \frac{k^2}{a^2} \delta_{c{\rm N}}
+\left( 1+a_2 \right) \frac{k^2}{a^2} \Psi
-\frac{k^2}{a^2}\frac{H}{\tp} 
\left[ a_2 \frac{\dot{\delta \phi}_{\rm N}}{H}
-(3\hat{c}_c^2+a_1+\epsilon_{\phi} a_2) \delta \phi_{\rm N} \right] 
\nonumber \\
& &
=-3\ddot{\Phi}-3\left( 2+a_1 \right) H \dot{\Phi}\,,
\label{delceq}
\\
& &
\ddot{\delta}_{b{\rm N}}+2H\dot{\delta}_{b{\rm N}}
+\frac{k^2}{a^2} \Psi
=-3\ddot{\Phi}-6 H \dot{\Phi}\,.
\label{delbeq}
\ea

The scalar perturbation equations derived in Sec.~\ref{scaeqsec} 
can be expressed in terms of the gauge-invariant gravitational 
potentials $\Psi$ and $\Phi$ as well as the other gauge-invariant 
perturbations given in Eq.~(\ref{delphiN}).
First of all, Eq.~(\ref{eqE2}) translates to 
\be
\Psi+\Phi=-\frac{H \alpha_{\rm M}}
{\dot{\phi}}\delta \phi_{\rm N}\,,
\label{quasi1}
\ee
where 
\be
\alpha_{\rm M}=\frac{\dot{q}_t}{Hq_t}\,.
\ee
\subsection{Quasi-static approximation for the modes 
deep inside the sound horizon}

In what follows, we employ the so-called quasi-static approximation 
for the modes deep inside the sound 
horizon \cite{Boisseau:2000pr,Tsujikawa:2007gd,DeFelice:2011hq}. 
Under this approximation scheme, the dominant contributions 
to the scalar perturbation equations of motion are the terms containing 
$\delta_{c{\rm N}}$, $\dot{\delta}_{c{\rm N}}$, 
$\delta_{b{\rm N}}$, and $k^2/a^2$. 
We also take into account the field mass squared $M_{\phi}^2$ appearing 
in Eq.~(\ref{calZeq}) to accommodate the case in which the field is heavy 
at high redshifts as in $f(R)$ models of 
late-time cosmic acceleration \cite{Hu:2007nk,Starobinsky:2007hu,Tsujikawa:2007xu}. 
The tachyonic instability can be avoided for $M_{\phi}^2 \geq 0$.
Applying the quasi-static approximation to 
Eqs.~(\ref{pereq1}) and (\ref{calZeq}), it follows that 
\ba
&&
\frac{k^2}{a^2} \left( 2q_t \Phi-D_6 \delta \phi_{\rm N} \right)
-(1+a_2)q_c\rho_c\delta_{c{\rm N}}
-\rho_b \delta_{b{\rm N}} \simeq 0\,,
\label{quasi2}\\
&&
M_{\phi}^2 \delta \phi_{\rm N}+\frac{k^2}{a^2} \left[
\frac{1}{2q_t}\left(q_s \hat{c}_s^2+D_6^2+\frac{4Hq_tD_6\alpha_{\rm M}}{\tp} \right)
\delta \phi_{\rm N}+D_6 \Psi-\frac{2H\alpha_{\rm M}q_t}{\dot{\phi}} 
\Phi \right]
\notag\\
&&
+\frac{Hq_c\rho_c}{\tp}\left[a_2\frac{\dot{\delta}_{c{\rm N}}}{H}
+\{3\hat{c}_c^2+a_1+a_2(a_1+\epsilon_{a_2})\}\delta_{c{\rm N}}\right]\simeq0\,,
\label{quasi3}
\ea
where we used Eq.~(\ref{csfinal}), and 
\be
\epsilon_{a_2}=\frac{\dot{a}_2}{Ha_2}\,.
\ee

The difference from the uncoupled Horndeski theories is that 
there exists the time derivative $\dot{\delta}_{c{\rm N}}$ 
in Eq.~(\ref{quasi3}). We solve Eqs.~(\ref{quasi1}), (\ref{quasi2}), 
and (\ref{quasi3}) for $\Psi$, $\Phi$, and $\delta \phi_{\rm N}$ 
to express them in terms of $\delta_{c{\rm N}}$, 
$\dot{\delta}_{c{\rm N}}$, and $\delta_{b{\rm N}}$. 
In doing so, we introduce the following dimensionless 
variables\footnote{The definition of $\Delta_2$ 
in Eq.~(6.14) of Ref.~\cite{Kase:2018aps} contains 
the term $q_t$ in the denominator, 
but the definition of $\Delta_2$ in Eq.~(\ref{Delta}) of this paper is 
different from the former in that $q_t$ is replaced by $q_t^2$ 
to make $\Delta_2$ dimensionless.},
\ba
&&
\Delta_1=\alpha_{\rm B}-\alpha_{\rm M}\,,\qquad 
\Delta_2=\frac{\dot{\phi}^2 q_s \hat{c}_s^2}{4H^2 q_t^2} 
\left( 1+\frac{2 q_t a^2 M_{\phi}^2 }{q_s \hat{c}_s^2 k^2} \right)\,,\qquad
\Delta_3=(1+a_2)\Delta_1+3\hat{c}_c^2+a_1+a_2(a_1+\epsilon_{a_2})\,,
\label{Delta} \nonumber \\
&&
\alpha_{\rm B}=-\frac{D_6 \dot{\phi}}{2 Hq_t}\,.
\ea
Then, $\Psi$, $\Phi$, and $\delta \phi_{\rm N}$ are expressed 
in the forms,
\ba
\hspace{-0.8cm}
\Psi &=& -\frac{a^2}{2q_t \Delta_2 k^2} \left[ 
\{\Delta_1\Delta_3+(1+a_2)\Delta_2\}q_c \rho_c \delta_{c{\rm N}}
+ (\Delta_1^2+\Delta_2)\rho_b \delta_{b{\rm N}}
+a_2q_c\Delta_1 \rho_c \frac{\dot{\delta}_{c{\rm N}}}{H}\right],
\label{Psif} \\
\hspace{-0.8cm}
\Phi &=& \frac{a^2}{2q_t \Delta_2 k^2} \left[ 
\{\alpha_{\rm B}\Delta_3+(1+a_2)\Delta_2\}q_c \rho_c \delta_{c{\rm N}}
+(\alpha_{\rm B}\Delta_1+\Delta_2) \rho_b \delta_{b{\rm N}}
+a_2q_c\alpha_{\rm B} \rho_c \frac{\dot{\delta}_{c{\rm N}}}{H}\right],\\
\delta \phi_{\rm N} &=&
-\frac{a^2 \dot{\phi}}{2H q_t \Delta_2 k^2} \left( 
\Delta_3q_c\rho_c \delta_{c{\rm N}}
+\Delta_1 \rho_b \delta_{b{\rm N}}
+a_2 q_c \rho_c \frac{\dot{\delta}_{c{\rm N}}}{H}
\right)\,.
\label{delphif} 
\ea
The gravitational slip parameter is given by
\be
\eta \equiv -\frac{\Phi}{\Psi}
=\frac{[\alpha_{\rm B}(\Delta_3+a_2 \epsilon_{\delta_c})+(1+a_2)\Delta_2]
q_c \Omega_c \delta_{c{\rm N}}+(\alpha_{\rm B}\Delta_1+\Delta_2)
\Omega_b \delta_{b{\rm N}}}
{[\Delta_1(\Delta_3+a_2\epsilon_{\delta_c})+(1+a_2)\Delta_2]
q_c \Omega_c \delta_{c{\rm N}}+(\Delta_1^2+\Delta_2)
\Omega_b \delta_{b{\rm N}}}\,,
\ee
where 
\be
\Omega_c=\frac{\rho_c}{3M_{\rm pl}^2H^2}\,,\qquad 
\Omega_b=\frac{\rho_b}{3M_{\rm pl}^2H^2}\,,\qquad 
\epsilon_{\delta_c}=\frac{\dot{\delta}_c}{H \delta_c}\,.
\ee
We substitute Eqs.~(\ref{Psif}), (\ref{delphif}), and the time derivative 
of Eq.~(\ref{delphif}) into Eqs.~(\ref{delceq}) and (\ref{delbeq}).
In doing so, the terms on the right hand sides of Eqs.~(\ref{delceq}) 
and (\ref{delbeq}) are neglected relative to those on their left hand sides. 
We also introduce the following dimensionless variables,
\be
Q_t=\frac{q_t}{M_{\rm pl}^2}\,,\qquad
\epsilon_{H}=\frac{\dot{H}}{H^2}\,,\qquad
\epsilon_{\Delta_i}=\frac{\dot{\Delta}_i}{H\Delta_i}\qquad ({\rm for}~i=1,2,3)\,,
\ee
and 
\be
b_1=\frac{3a_2^2q_c\Omega_c}{2Q_t\Delta_2}\,,\qquad
b_2=-b_1(1+2\epsilon_H+\alpha_{\rm M}+\epsilon_{\Delta_2}-2a_1-2\epsilon_{a_2})\,,\qquad 
b_3=\frac{3a_2\Delta_1\Omega_b}{4Q_t\Delta_2}\,,
\ee
where $b_1>0$ under the absence of ghosts and Laplacian/tachyonic instabilities.
Then, it follows that 
\ba
&&
\ddot{\delta}_{c{\rm N}}+\frac{2+a_1+b_2}{1+b_1}H\dot{\delta}_{c{\rm N}}
+\frac{2b_3}{1+b_1} H\dot{\delta}_{b{\rm N}}+\frac{\hat{c}_c^2}{1+b_1} 
\frac{k^2}{a^2}\delta_{c{\rm N}}
-\frac{3H^2}{2G} \left(G_{cc}\Omega_c\delta_{c{\rm N}}
+G_{cb}\Omega_b\delta_{b{\rm N}}\right)
\simeq0\,,\label{delceq2} \\
&&
\ddot{\delta}_{b{\rm N}}+2H\dot{\delta}_{b{\rm N}}
-2Hq_cb_3\frac{\Omega_c}{\Omega_b}\dot{\delta}_{c{\rm N}}
-\frac{3H^2}{2G}\left(G_{bc}\Omega_c\delta_{c{\rm N}}
+G_{bb}\Omega_b\delta_{b{\rm N}}\right)\simeq0\,, 
\label{delbeq2}
\ea
where $G=1/(8\pi M_{\rm pl}^2)$, and 
the effective gravitational couplings are given by  
\ba
\hspace{-0.7cm}
G_{cc} &=& 
\frac{2q_c}{2Q_t\Delta_2+3a_2^2q_c\Omega_c}
\left[(1+a_2)^2\Delta_2+\Delta_3\left\{\Delta_1+3\hat{c}_c^2
+a_1+a_2\left(1+\epsilon_H+\alpha_{\rm M}+\Delta_1
-a_1+\epsilon_{\Delta_2}-\epsilon_{\Delta_3}\right)\right\}\right]G\,,
\label{Gcc} \\
\hspace{-0.7cm}
G_{cb} &=&\frac{2}{2Q_t\Delta_2+3a_2^2q_c\Omega_c}
\left[(1+a_2)(\Delta_1^2+\Delta_2)+\Delta_1\left\{3\hat{c}_c^2+a_1
+a_2(1+\alpha_{\rm M}+\epsilon_H-\epsilon_{\Delta_1}+\epsilon_{\Delta_2})
\right\}\right]G\,,\\
\hspace{-0.7cm}
G_{bc} &=& 
\frac{q_c\left[\Delta_1\Delta_3+(1+a_2)\Delta_2\right]}{Q_t\Delta_2}G\,,\\
\hspace{-0.7cm}
G_{bb} &=& \frac{\Delta_1^2+\Delta_2}{Q_t\Delta_2}G\,. 
\label{Gbb}
\ea
The gravitational couplings $G_{cc}$ and $G_{cb}$, which are multiplied by 
$\Omega_c\delta_{c{\rm N}}$ and $\Omega_b\delta_{b{\rm N}}$ respectively, 
affect the growth of CDM density contrast $\delta_{c{\rm N}}$. 
Meanwhile, the baryon density contrast $\delta_{b{\rm N}}$ has 
the gravitational couplings multiplied by 
$\Omega_b\delta_{b{\rm N}}$ and $\Omega_c\delta_{c{\rm N}}$, respectively. 
The baryon perturbation is affected by the evolution of the CDM perturbation, 
and vice versa. 

In the absence of matter couplings ($f_1=0=f_2$), the quantities 
$a_1,a_2,b_1,b_2,b_3$ vanish. In this case we have $q_c=1$, $\hat{c}_c^2=0$ 
and $\Delta_3=\Delta_1$, so that 
$G_{cc},G_{cb},G_{bc}$ reduce to the same form as $G_{bb}$. 
We note that $G_{bb}$ is equivalent to the effective 
gravitational coupling $G_{\rm eff}$ derived for uncoupled 
Horndeski theories \cite{DeFelice:2011hq,Kase:2018aps}. 
When the matter couplings $f_1$ and $f_2$ are present, 
$G_{cc},G_{cb},G_{bc}$ are generally different from 
$G_{bb}$. 

Apart from the specific coupling having the dependence $f_2 \propto n_c^{-1}$, 
$\hat{c}_c^2$ is different from 0.  
For $\hat{c}_c^2>0$, the growth of $\delta_{c{\rm N}}$ is prevented by 
the positive pressure induced by the coupling $f_2(n_c)$, 
whereas, for $\hat{c}_c^2<0$, there is the additional enhancement of 
$\delta_{c{\rm N}}$ besides the growth induced by gravitational instabilities.
For the consistency with the observed galaxy power spectrum in the linear 
regime of perturbations, the effective sound speed squared needs to be 
much smaller than 1, 
say, $|\hat{c}_c^2| \lesssim 10^{-5}$ in unified perfect fluid models 
of dark energy and dark matter \cite{Waga}. 
We note that the $\phi$ and $X$ dependence in $f_1$ and $f_2$ 
does not modify the value of $\hat{c}_c^2$. 

The above results have been obtained for the gauge-invariant density contrasts 
(\ref{delI}) for CDM and baryons. 
There is also the gauge-invariant CDM density perturbation (\ref{delrhohat}) 
containing the effect of coupling $f_1$. Dividing $\widehat{\delta \rho}_{c{\rm N}}$ 
by the background total CDM density $\hat{\rho}_c=(1+f_1)\rho_c$, 
the corresponding density contrast is 
\be
\hat{\delta}_{c{\rm N}} \equiv 
\frac{\widehat{\delta \rho}_{c{\rm N}}}{\hat{\rho}_c}
=\delta_{c{\rm N}}+\frac{f_{1,\phi} \delta \phi_{\rm N}
+f_{1,X} \dot{\phi} (\dot{\delta \phi}_{\rm N}
-\dot{\phi}\Psi)}{1+f_1}\,.
\label{delcN}
\ee
Under the quasi-static approximation, 
the terms containing $\delta \phi_{\rm N}$, $\dot{\delta \phi}_{\rm N}$, and 
$\Psi$ in Eq.~(\ref{delcN}) are suppressed relative to $\delta_{c{\rm N}}$. 
Indeed, this property can be confirmed by the solutions 
(\ref{Psif}) and (\ref{delphif}) in the small-scale limit. 
In this case, we have $\hat{\delta}_{c{\rm N}} \simeq \delta_{c{\rm N}}$ 
and hence the effective gravitational couplings for $\hat{\delta}_{c{\rm N}}$ 
are approximately the same as those derived above for the perturbations 
deep inside the sound horizon.

\subsection{Concrete theories}

On using the general formulas (\ref{Gcc})-(\ref{Gbb}), we compute the effective 
gravitational couplings for specific theories and the choice of couplings.
We consider three different cases: 
(i) $f_1=f_1(\phi)$, $f_2=0$, 
(ii) $f_1=f_1(\phi)$, $f_2=f_2(n_c,\phi)$, and 
(iii) $f_1=0$, $f_2=f_2(n_c,X)$.

\subsubsection{Theories with $f_1=f_1(\phi)$ and $f_2=0$}

In this case, we introduce the dimensionless coupling 
\be
Q=\frac{M_{\rm pl} f_{1,\phi}}{1+f_1}\,. 
\label{Q}
\ee
Then, the right hand sides of Eqs.~(\ref{rhoccon}) and (\ref{rhodecon}) 
reduce to $Q \hat{\rho}_c \dot{\phi}/M_{\rm pl}$ 
and $-Q \hat{\rho}_c \dot{\phi}/M_{\rm pl}$, respectively, showing that 
the quantity $Q$ characterizes the strength of nonminimal coupling $f_1(\phi)$. 
The quantities $a_1$, $a_2$, $q_c$, $\hat{c}_c^2$, and 
$\Delta_3$ are given, respectively, by 
\be
a_1=Qx,\qquad a_2=0\,,\qquad q_c=1+f_1\,,\qquad 
\hat{c}_c^2=0\,,\qquad
\Delta_3=\Delta_1+Qx\,,
\ee
where 
\be
x \equiv \frac{\dot{\phi}}{M_{\rm pl} H}\,.
\label{xdef}
\ee
Since the coupling $f_2$ is absent, $\hat{c}_c^2$ vanishes.
{}From Eqs.~(\ref{Gcc})-(\ref{Gbb}), it follows that 
\ba
& &
G_{cc}=\frac{G}{Q_t} \left( 1+f_1 \right) \left[ 
1+\frac{(Qx+\Delta_1)^2}{\Delta_2} \right]\,,\qquad 
G_{cb}=\frac{G}{Q_t} \left[1+\frac{\Delta_1 
(Qx+\Delta_1)}{\Delta_2} \right]\,,\nonumber \\
& &
G_{bc}=\frac{G}{Q_t} \left( 1+f_1 \right) \left[1+\frac{\Delta_1 
(Qx+\Delta_1)}{\Delta_2} \right]\,,\qquad 
G_{bb}=\frac{G}{Q_t} \left( 1+ \frac{\Delta_1^2}{\Delta_2} 
\right)\,.
\label{Gcon1}
\ea
As long as the stability conditions of tensor and scalar 
perturbations are satisfied, it follows that 
$Q_t>0$, $q_c=1+f_1>0$, and $\Delta_2>0$. 
Hence both $G_{cc}$ and $G_{bb}$ are positive, with the positive 
scalar-matter interactions characterized by 
$(Qx+\Delta_1)^2/\Delta_2>0$ and 
$\Delta_1^2/\Delta_2>0$. 
According to our knowledge, the gravitational couplings 
(\ref{Gcon1}) have not been explicitly derived 
in coupled Horndeski theories given by the action (\ref{Sg}).

In the following, we specify the scalar-graviton sector to 
the minimally coupled k-essence given by the action,
\be
{\cal S}_{\rH}=
\int {\rm d}^4 x \sqrt{-g} 
\left[ \frac{M_{\rm pl}^2}{2} R+G_2(\phi,X)
\right]\,.
\label{Kess}
\ee
We also consider a light-mass scalar field satisfying 
\be
\frac{q_t a^2 M_{\phi}^2}{q_s \hat{c}_s^2 k^2} \ll 1\,,
\label{massless}
\ee
in $\Delta_2$ of Eq.~(\ref{Delta}). 
We also adopt the condition (\ref{massless}) for the theories 
discussed later in Secs.~\ref{model2} and \ref{model3sec}.
Then, we have 
\be
Q_t=1\,,\qquad 
q_s \hat{c}_s^2=2M_{\rm pl}^2 G_{2,X}\,,\qquad 
\alpha_{\rm M}=0\,,\qquad 
\alpha_{\rm B}=0\,,\qquad
\Delta_1=0\,,\qquad \Delta_2=\frac{x^2}{2}G_{2,X}\,.
\ee
{}From Eq.~(\ref{Gcon1}), we obtain 
\be
G_{cc}=G \left( 1+f_1 \right) \left( 1+
\frac{2Q^2}{G_{2,X}} \right)\,,\qquad 
G_{cb}=G\,,\qquad G_{bc}=G(1+f_1)\,,\qquad 
G_{bb}=G\,.
\label{Gcon}
\ee
The coupling term $2Q^2/G_{2,X}$ in $G_{cc}$ coincides 
with that derived in Refs.~\cite{Amendola:2004qb,Tsujikawa:2007gd}. 
For a canonical scalar field given by the Lagrangian $G_2=X-V(\phi)$, 
the scalar-matter interaction in $G_{cc}$ reduces to the well-known 
form $2Q^2$ \cite{Amendola:2003wa,Ade15}, 
which enhances the growth rate of CDM perturbations.

\subsubsection{Theories with $f_1=f_1(\phi)$ and $f_2=f_2(n_c,\phi)$}
\label{model2}

Besides the nonminimal coupling $f_1=f_1(\phi)$, we consider the case 
in which the second interaction of the form $f_2=f_2(n_c,\phi)$ is present.
For concreteness, we focus on the couplings, 
\be
f_1=f_1(\phi)\,,\qquad f_2=c n_c^{-1}+F_2(\phi)\,,
\ee
where $f_1$, $F_2$ are functions of $\phi$, and 
$c$ is a constant. 
In this case, we have $\hat{c}_c^2=0$.
Together with the coupling $Q$ defined in Eq.~(\ref{Q}),
we introduce the following dimensionless quantities, 
\be
q=\frac{M_{\rm pl} Q_{,\phi}}{Q}\,,\qquad 
\Omega_{f_2}=\frac{c \dot{\phi}}{3M_{\rm pl}^2 H^2}\,,
\ee
and consider a light scalar field whose mass $M_{\phi}$ 
is in the range (\ref{massless}).

For the scalar-graviton sector, we adopt a canonical scalar field 
given by the action 
\be
{\cal S}_{\rH}=
\int {\rm d}^4 x \sqrt{-g} 
\left[ \frac{M_{\rm pl}^2}{2} R+X-V(\phi)
\right]\,,
\ee
where $V(\phi)$ is the field potential.
Then, it follows that 
\be
a_1=\frac{(1+f_1) \Omega_c Qx+(3+\epsilon_{\phi})
\Omega_{f_2}}{(1+f_1) \Omega_c+\Omega_{f_2}}\,,
\qquad a_2=-\frac{\Omega_{f_2}}{(1+f_1) \Omega_c+\Omega_{f_2} }\,,
\qquad q_c=1+f_1+\frac{\Omega_{f_2}}{\Omega_c}\,,
\ee
and 
\ba
& &
\alpha_{\rm M}=0\,,\qquad \alpha_{\rm B}=0\,,\qquad 
\Delta_1=0\,,\qquad \Delta_2=\frac{x^2}{2}\,,\qquad 
\Delta_3=\frac{(1+f_1) \Omega_c Qx}{(1+f_1) \Omega_c+\Omega_{f_2}}\,, 
\nonumber \\
& &
\epsilon_{\Delta_1}=0\,,\qquad 
\epsilon_{\Delta_2}=2 \left( \epsilon_{\phi}-\epsilon_H \right)\,,\qquad 
\epsilon_{\Delta_3}=\frac{(1+f_1) \Omega_c (\epsilon_{\phi}-\epsilon_H
+qx)+\Omega_{f_2}[(Q+q)x-3-\epsilon_H]}
{(1+f_1) \Omega_c+\Omega_{f_2}}\,.
\ea
The quantity $x$, which is defined by Eq.~(\ref{xdef}), is expressed as 
$x=\epsilon \sqrt{2\Delta_2}$, where $\epsilon =+1$ for 
$\dot{\phi}>0$ and  $\epsilon =-1$ for $\dot{\phi}<0$.
The effective gravitational couplings (\ref{Gcc})-(\ref{Gbb}) reduce, 
respectively, to 
\ba
& &
G_{cc}=G \left( 1+f_1 \right) \left[ 1+2Q^2 
-\Omega_{f_2}\frac{2\Delta_2 (1-2Qq)-4 \epsilon Q \sqrt{2\Delta_2}
+3\Omega_{f_2} (1+2Q^2)}
{2 \Omega_c \Delta_2 (1+f_1)
+\Omega_{f_2} (2\Delta_2+3\Omega_{f_2})} \right]\,,\nonumber \\
& &
G_{cb}=G \frac{2\Omega_c (1+f_1) \Delta_2}
{2[\Omega_c (1+f_1)+\Omega_{f_2}] \Delta_2
+3\Omega_{f_2}^2}\,,\qquad 
G_{bc}=G  \left( 1+f_1 \right)\,,\qquad 
G_{bb}=G\,.
\label{Gf}
\ea
Compared to the values given in Eq.~(\ref{Gcon}), 
the $n_c$ dependence in $f_2$ modifies both $G_{cc}$ and $G_{cb}$. 
We note that there is no modification to the  gravitational couplings 
from the term $F_2(\phi)$ in $f_2$.

If we consider the theories with \cite{Boehmer:2015sha}
\be
f_1=0\,,
\ee
i.e., $Q=0$, Eq.~(\ref{Gf}) gives 
\be
G_{cc}=G_{cb}=G \frac{2\Omega_c \Delta_2}
{2(\Omega_c+\Omega_{f_2})\Delta_2+3\Omega_{f_2}^2}\,,
\qquad G_{bc}=G_{bb}=G\,,
\label{Gf2}
\ee
where $G_{cc}$ matches with that derived in Eq.~\cite{Koivisto:2015qua}.
Since $\Delta_2>0$ and $\Omega_c>0$, the gravitational couplings  
with CDM can be smaller than $G$ (i.e., $G_{cc}=G_{cb}<G$) for
\be
\Omega_{f_2}>0\,,\quad {\rm i.e.,} \quad 
c \dot{\phi}>0\,.
\ee
This is an explicit example in which the $n_c$ dependence in 
$f_2$ allows the possibility for realizing weak gravity on scales 
relevant to the linear growth of large-scale structures. 
We note that, under the absence of ghosts and Laplacian instabilities,  
the effective gravitational coupling in uncoupled Horndeski theories is 
destined to be larger than $G$ \cite{GW5,Kase:2018aps}. 
Hence, the coupled model with weak gravitational interactions 
can be observationally distinguished from uncoupled Horndeski theories.  
It is of interest to explore further whether 
the coupled dark energy model with the coupling $f_2 \propto n_c^{-1}$ 
can be in better fit to the observational data in comparison to the 
$\Lambda$CDM model. 
In the presence of nonvanishing $f_1(\phi)$, 
the gravitational couplings $G_{cc}$ and $G_{cb}$ in Eq.~(\ref{Gf}) 
differ from those in Eq.~(\ref{Gf2}),  
so the observational signatures are different.

\subsubsection{Theories with $f_1=0$ and $f_2=f_2(n_c,X)$}
\label{model3sec}

Finally, we study the effect of $X$-dependence in $f_2$ on the 
effective gravitational couplings. 
For concreteness, let us consider the functions,
\be
f_1=0\,,\qquad 
f_2=\beta n_c^{-1} X^s\,, 
\label{model3}
\ee
where $\beta$ and $s$ are constants. 
For this choice of $f_2$, 
$\hat{c}_c^2$ vanishes. On using the definition of fluid 
four-velocity (\ref{udef}), the interacting action 
(\ref{Sint}) can be written as 
\be
{\cal S}_{\rm int} = 
\int {\rm d}^4x\sqrt{-g}\,\beta X^s u_c^{\mu} \partial_{\mu} \phi\,.
\label{Sint3}
\ee
On the flat FLRW background (\ref{metric}) with $N=1$, 
the term $u_c^{\mu} \partial_{\mu} \phi$ is equivalent to $\dot{\phi}$. 
Then, the above interacting Lagrangian is proportional to $\dot{\phi}^{2s+1}$.
For the scalar-graviton sector, we adopt the minimally coupled k-essence 
described by the action (\ref{Kess}). Then, Eqs.~(\ref{rhode}) and (\ref{Pde}) 
reduce, respectively, to 
\ba
\rho_{\rm DE}&=&-G_2+\dot{\phi}^2G_{2,X}+2s\bar{\beta}\dot{\phi}^{2s+1}
\,,\label{rhode3}\\
P_{\rm DE}&=&G_2+\bar{\beta}\dot{\phi}^{2s+1}\,,
\label{Pde3}
\ea
where $\bar{\beta}=2^{-s}\beta$. 
For the power, 
\be
s=\frac{1}{2}\,, 
\label{scaling}
\ee
the last terms in Eqs.~(\ref{rhode3}) and (\ref{Pde3}) are proportional to 
$\dot{\phi}^2$, i.e., the same time dependence as the standard kinetic 
term $X$ in $G_2$. 
If $s\neq1/2$, the last terms in Eqs.~(\ref{rhode3}) and (\ref{Pde3}) 
can dominate over $X$ either in the asymptotic past or future. 
In the following, we focus on the model with the power (\ref{scaling}). 
Then, it follows that 
\be
a_1=\frac{\bar{\beta}x^2(3+2\epsilon_{\phi})}{3\Omega_c+\bar{\beta}x^2}\,,\qquad
a_2=-\frac{\bar{\beta}x^2}{3\Omega_c+\bar{\beta}x^2}\,,\qquad
q_c=1+\frac{\bar{\beta}x^2}{3\Omega_c}\,,\qquad
\hat{c}_c^2=0\,,
\label{nodim1}
\ee
where $x$ is defined by Eq.~(\ref{xdef}), and 
\ba
& &
Q_t=1\,,\qquad 
q_s \hat{c}_s^2=2M_{\rm pl}^2(G_{2,X}+\bar{\beta})\,\qquad
\alpha_{\rm M}=0\,,\qquad 
\alpha_{\rm B}=0\,,\qquad 
\nonumber \\
& &
\Delta_1=0\,,\qquad 
\Delta_2=\frac{x^2}{2}(G_{2,X}+\bar{\beta})\,,\qquad 
\Delta_3=0\,.
\label{nodim2}
\ea
Substituting Eqs.~(\ref{nodim1}) and (\ref{nodim2}) into (\ref{Gcc})-(\ref{Gbb}), 
we obtain
\ba 
& &
G_{cc}=G_{cb}=
G\left( 1+\frac{G_{2,X}+2\bar{\beta}}{G_{2,X}+\bar{\beta}}
\frac{\bar{\beta}x^2}{3\Omega_c} \right)^{-1}\,,\label{Gs3}\\
& &
G_{bc}=G_{bb}=1\,.
\ea
From Eq.~(\ref{nodim1}), the ghost does not appear from the CDM sector 
for $\bar{\beta}>0$. Moreover, the condition $G_{2,X}+\bar{\beta}>0$ 
is required for the positivity of $q_s \hat{c}_s^2$ in Eq.~(\ref{nodim2}). 
Then, the second term in the parenthesis of Eq.~(\ref{Gs3}) is 
positive, so that $G_{cc}=G_{cb}<G$.
This is the explicit example in which the $X$-dependence in $f_2$ leads to 
the realization of weak gravity on cosmological scales.

The interacting action (\ref{Sint3}) with the power (\ref{scaling}) 
can be further generalized to the theories containing the 
non-linear dependence of $u_c^{\mu}\partial_{\mu}\phi$, e.g., 
\be
{\cal S}_{\rm int} = \int {\rm d}^4x\sqrt{-g}\,\beta X^s (u_c^{\mu} \partial_{\mu} \phi)^{2-2s}\,. 
\label{model4}
\ee
By setting $s=0$, the interacting action (\ref{model4}) recovers the theory studied 
in Ref.~\cite{Pourtsidou:2016ico} as a special case. The detailed analysis of 
cosmological dynamics for the extended coupling (\ref{model4}) is 
given in Ref.~\cite{Kase:2019mox} (which was submitted one month 
after the initial submission of this paper).

\section{Conclusions}
\label{consec}

We studied the interacting dark energy scenario in which a scalar field $\phi$ 
is coupled to the CDM perfect fluid given by the Schutz-Sorkin action (\ref{SM}). 
The scalar-graviton sector is described by the Horndeski action (\ref{Sg}) 
with the tensor propagation speed squared equivalent to that of light. 
We considered the new interacting action (\ref{Sint}) containing the 
$X$ dependence in the couplings $f_1$ and $f_2$.
Our analysis is sufficiently general in that it accommodates a wide 
variety of nonminimal and derivative couplings studied 
in the literature \cite{Wetterich,Amendola99,Boehmer:2015kta,Boehmer:2015sha,Koivisto:2015qua,Barros}.
Moreover, unlike most of past related papers, we did not 
restrict the dark energy field to quintessence or k-essence.

In Sec.~\ref{eomsec}, we derived the background equations of motion 
on the flat FLRW background in the forms (\ref{Eq00})-(\ref{Eqphi}).
As long as the quantity $q_s$ defined by Eq.~(\ref{qs}) does not vanish, 
the dynamical system can be solved for $\dot{H}$ and $\ddot{\phi}$.
Indeed, the stability analysis performed in Sec.~\ref{stasec} 
leads to the condition $q_s>0$ to avoid the ghost associated with the scalar
perturbation. We also identified the total CDM density $\hat{\rho}_c$ 
and pressure $\hat{P}_c$ containing the effect of interactions 
with the scalar field, as Eqs.~(\ref{trhoc}) and (\ref{Prhoc}). 
We showed that CDM interacts with 
dark energy according to Eqs.~(\ref{rhoccon}) and 
(\ref{rhodecon}), whose signs on the right hand sides are 
opposite to each other. 

In Sec.~\ref{actionsec}, we expanded the action up to second order 
in scalar perturbations without fixing any gauge conditions.
We explicitly computed the quadratic-order actions arising from 
${\cal S}_{\rm H}$, ${\cal S}_{M}$, ${\cal S}_{\rm int}$ and finally 
took the sum of them on account of the background equations.
The final second-order action ${\cal S}_s^{(2)}$ is of the form 
(\ref{faction}), where $L_{\rm int}$ is the Lagrangian arising from 
the couplings $f_1$ and $f_2$. In Sec.~\ref{scaeqsec}, we also 
obtained the full linear perturbation equations of motion in 
the gauge-ready form, i.e., ready for choosing any 
particular gauges.

In Sec.~\ref{stasec}, we first introduced a number of gauge-invariant 
perturbed quantities and discussed several different gauge choices. 
In the flat and unitary gauges, we derived the second-order actions 
of dynamical scalar perturbations after eliminating nondynamical 
quantities and identified stability conditions in the small-scale limit. 
The conditions for the absence of ghosts and Laplacian instabilities 
are independent of the gauge choices. 
The ghosts are absent under the conditions (\ref{qscon}) and 
(\ref{CDMcon}). For CDM satisfying $\hat{c}_c^2=0$, 
the off-diagonal components of matrix ${\bm B}$ do not 
modify the CDM sound speed, i.e., $c_{{\rm CDM}}^2=0$, but 
they affect the propagation speed squared $c_s^2$ 
of the scalar degree of freedom. We showed that $c_s^2$
is the sum of $\hat{c}_s^2$ and $\Delta c_s^2$, whose explicit 
expressions are given, respectively, by 
Eqs.~(\ref{csfinal}) and (\ref{csfinalde}).

In Sec.~\ref{Geffsec}, we derived the effective gravitational couplings 
for CDM and baryon density perturbations by using the quasi-static 
approximation for the modes deep inside the sound horizon. 
We obtained the Bardeen gravitational potentials $\Psi$, $\Phi$, and 
the field perturbation $\delta \phi_{\rm N}$ 
in the gauge-independent manner.
The difference from the uncoupled case is that the time derivative 
$\dot{\delta}_{c{\rm N}}$ appears in the expressions of 
$\Psi$, $\Phi$, $\delta \phi_{\rm N}$ given by
Eqs.~(\ref{Psif})-(\ref{delphif}). 
Taking the time derivative of $\delta \phi_{\rm N}$ in  
Eq.~(\ref{delphif}) gives rise to the second derivative $\ddot{\delta}_{c{\rm N}}$, 
so we need to solve Eq.~(\ref{delceq}) for $\ddot{\delta}_{c{\rm N}}$ 
to obtain the closed-form equation for $\delta_{c{\rm N}}$.
As a result, the CDM and baryon density contrasts obey Eqs.~(\ref{delceq2}) 
and (\ref{delbeq2}), respectively, with the effective gravitational 
couplings (\ref{Gcc})-(\ref{Gbb}). 

We applied our general formulas of $G_{cc}$, $G_{cb}$, $G_{bc}$, 
$G_{bb}$ to three different forms of couplings and discussed how 
they reproduce the effective gravitational couplings known 
in the literature. In particular, the $n_c$ or $X$ dependence 
in $f_2$ offers an interesting possibility for realizing $G_{cc}$ and 
$G_{cb}$ smaller than $G$. 
In uncoupled Horndeski theories, the gravitational couplings 
are usually larger than $G$.
Moreover, the presence of matter couplings $f_1$ and $f_2$ gives 
rise to the values of $G_{cc}$, $G_{cb}$, $G_{bc}$, $G_{bb}$ generally different from 
each other, while this is not the case in uncoupled Horndeski theories.  
These properties show that the coupled theories with weak gravitational interactions 
can be observationally distinguished from uncoupled theories.

In this paper we did not construct particular models of coupled dark energy, 
but it is of interest to do so to observationally probe the signature of 
interactions with CDM. 
First of all, theoretically consistent models need to satisfy all the small-scale stability conditions derived in Sec.~\ref{stasec}. 
The next step is to predict observational signatures of models 
both at the background and perturbation levels, e.g., 
the dark energy and CDM equations of state and the growth 
of perturbations. Then, the models should be confronted with 
numerous observational data associated with the cosmic expansion and 
growth histories. These interesting issues are left for future works.

\section*{Acknowledgements}

RK is supported by the Grant-in-Aid for Young Scientists B 
of the JSPS No.\,17K14297. 
ST is supported by the Grant-in-Aid for Scientific Research Fund of the JSPS No.\,19K03854 and
MEXT KAKENHI Grant-in-Aid for Scientific Research on Innovative Areas
``Cosmic Acceleration'' (No.\,15H05890).


\end{document}